\documentclass[aps,prd,twocolumn,showpacs]{revtex4}
\usepackage{color}
\usepackage{graphicx}
\usepackage{epsfig}
\usepackage{bm,hyperref}
\usepackage[utf8x]{inputenc}

\def\nn {\nonumber}

\newcommand{\be}{\begin{equation}}
\newcommand{\ee}{\end{equation}}
\newcommand{\bea}{\begin{eqnarray}}
\newcommand{\eea}{\end{eqnarray}}

\newcommand{\ep}{\epsilon}

\newcommand{\om}{\omega}

\newcommand{\Ds}{D \!\!\! /}

\newcommand{\bp}{\boldsymbol{p}}

\newcommand{\del}{\partial}

\newcommand{\munu}{{\mu\nu}}

\newcommand{\FB}[1]{\left(#1\right)}

\begin{document}
TIFR/TH/19-1

\title{Impact of different extended components of mean field models on transport coefficients of quark matter
and their causal aspects}
\author{Chowdhury Aminul Islam}
\email{chowdhury.aminulislam@gmail.com}
\affiliation{Department of theoretical Physics, Tata Institute of fundamental Research, 
Homi Bhabha Road, Mumbai 400005, India}
\author{Jayanta Dey}
\email{jayantad@iitbhilai.ac.in}
\affiliation{Indian Institute of Technology Bhilai, GEC Campus, Sejbahar, Raipur 492015, Chhattisgarh, India}
\author{Sabyasachi Ghosh}
\email{sabyaphy@gmail.com}
\affiliation{Indian Institute of Technology Bhilai, GEC Campus, Sejbahar, Raipur 492015, Chhattisgarh, India}
%

\begin{abstract}
Role of different extensions of Nambu\textendash Jona-Lasinio (NJL) model like addition 
of vector interaction, Polyakov loop extended version (PNJL) and the entangled 
PNJL (EPNJL) models on transport coefficients like shear viscosity, bulk viscosity, electrical conductivity 
and thermal conductivity are critically analyzed. We have considered the standard 
expressions of transport coefficients, obtained in relaxation time approximation
of kinetic theory. Influence of temperature dependent order parameters on temperature
profile of transport coefficients are analyzed. Causal aspect of massless case to these different extended components of 
mean field models are also picturized, where an approximated lower and upper bound are
drawn for shear relaxation time.


%
\end{abstract}

\maketitle
\section{Introduction}
Microscopic calculations of transport coefficients for highly dense quark matter, which may be
seen in astrophysical object like compact stars,
are an important  input  in  modeling  an  array  of  astrophysical
phenomena. Refs.~\cite{Astro1,Astro2,Astro3,Astro4} have gone through
these microscopic estimations. 
Future experimental facilities, such as Facility for Antiproton and Ion Research at GSI, Germany~\cite{FAIR} and the NICA at JINR, Russia~\cite{NICA}
aim to probe similar kind of high density zone in their laboratories.
Transport coefficients of highly dense matter, produced there, may have 
influence on different phenomenological quantities like spectra, flow,
which can be constructed from experimental data, measured by their detector set up.

On the other hand, a baryon free hot system can also be a matter of
interest to know its transport coefficients values. It is believed that
our early universe went through this state, just after few micro second
from big-bang. RHIC experiments at BNL, USA and LHC experiments at CERN, Switzerland
had reached this high
temperature and baryon free zone and their experimental
data~\cite{RHIC1,RHIC2,LHC1,LHC2,LHC3,LHC4}
indicate that the matter almost behave like a nearly perfect fluid.
A very small values of shear viscosity to entropy density ratio $\eta/s$
corresponds to this nature and this small values of $\eta/s$ has been searched
as input guess values in viscous hydrodynamic model analysis during the matching
experimental data of elliptic flow~\cite{hydro_rev,hydro_rev2,hydro_rev3}.
This small value of $\eta/s$ from experimental side throws a challenge to
the theoretical side, where microscopic calculations of $\eta/s$ for quark
matter can be done. Estimated values of $\eta/s$ from perturbative quantum chromodynamics (pQCD)
at leading order~\cite{pQCD1,pQCD2} are found to be quite larger than its experimental value.
However, Ref.~\cite{pQCD3} has recently found a significant drop of this value in next-to-leading 
order calculation but at the end of the article, the possibility of non-perturbative components 
in $\eta/s$ has not been ruled out. The non-perturbative temperature domain of QCD can be
well mimicked by effective QCD model calculations like Nambu\textendash\,Jona-Lasinio (NJL)
model and quark-meson (QM) models. In Refs~\cite{G_IFT,G_IFT2,G_CAPSS,Weise2,LKW,klevansky,klevansky2,Redlich_NPA,HMPQM1,HMPQM2,Marty,Deb,Kinkar_PNJL}, this type microscopic calculation of shear viscosity via
different effective QCD models has been performed. Among them, 
Refs.~\cite{G_IFT,G_IFT2,G_CAPSS,Weise2,LKW,klevansky,klevansky2,Redlich_NPA,Marty,Deb} have adopted
NJL model, Ref.~\cite{Kinkar_PNJL} has further incorporated the background gauge field through its Polyakov extended version.
There are many possible additional sources by which NJL model can be modified into
different versions. For example, addition of vector interaction, Polyakov loop extension, entangled Polyakov
loop extensions can modify the NJL model structure.
In present article, we have tried to investigate the impact of the different additional sources of NJL model
on $\eta/s$ calculations as well as for other transport coefficients like bulk viscosity, electrical conductivity and thermal conductivity. 

The article is organized as follows. In Sec.~\ref{sec:model}, formalism part of different versions of 
NJL model has been briefly addressed and in Sec.~\ref{sec:tr}, the expressions of different
transport coefficients are derived in kinetic theory framework along with their causal extensions. Then, Sec.~\ref{sec:Res}
has provided the detail numerical discussion, which have explored the impact of different extensions of NJL 
model on transport coefficient and at last, we have summarized our studies.

\section{Formalism of NJL model with different extensions}
\label{sec:model}
In this section we briefly discuss the mean field models that 
we have employed in our work. First we talk about the NJL model 
for two flavor case and introduce vector interaction in the picture. 
Then we extend it by introducing the Polyakov 
loop field known as PNJL model, through which the deconfinement dynamics can be
mimicked. In PNJL model the correlation between the chiral and deconfinement 
dynamics is weak. We impose a strong correlation between these two through 
Polyakov loop dependent coupling constants \textendash\, this is known as 
entangled PNJL (EPNJL) model.

\subsection{NJL}
\label{ssec:njl}
\begin{figure} [ht]
 \includegraphics[scale=0.8]{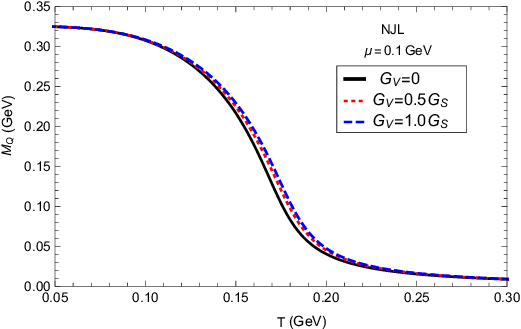}
 \caption{Temperature dependence of constituent quark masses for 
 $\frac{G_V}{G_S}=0$ (solid line), $0.5$ (dotted line), $1$ (dashed line) at $\mu=0.1$ GeV.}
 \label{fig:mass}
 \end{figure}

Let us start with NJL model first.
Here we are interested in two light quark flavors and we also include the 
isoscalar vector interaction which plays crucial role specially for system with 
finite density. The Lagrangian 
is~\cite{Kunihiro:1991qu,Klevansky:1992qe,Hatsuda:1994pi,Buballa:2003qv}:
\bea
&&\mathcal{L}_{\rm{NJL}} = \bar{\psi}(i\gamma_{\mu}\partial^{\mu}-m_0+\gamma_0\mu)\psi
+ \frac{G_{S}}{2}[(\bar{\psi}\psi)^{2}
\nn\\
&&~~~~~~~+(\bar{\psi}i\gamma_{5}\vec{\tau}\psi)^{2}]
- \frac{G_{V}}{2}(\bar{\psi}\gamma_{\mu}\psi)^2 
\label{eqn:njl},
\eea
where, $m_0 = m_0\times\bf{1}$, with $\bf{1}$ being the identity matrix and $m_{u}=m_{d}=m_0$; 
$\mu$ is the chemical potential; $\vec\tau$ is Pauli matrix; $G_S$ and $G_V$ are the four scalar 
and isoscalar-vector type coupling constants, respectively. The value of $G_V$ is not fixed 
through parameter fitting, rather it is used as a free parameter which can take values within 
the range $0\leq{G_V}/{G_S}\leq 1$. With the inclusion of vector interaction we now have another 
condensate as quark number density $n=\langle \bar{\psi}\gamma^0 \psi\rangle$~\cite{Buballa:2003qv,Kashiwa:2006rc} 
along with the usual chiral condensate $\Sigma=\langle \bar{\psi}\psi\rangle$. 
Chiral condensate will build the link between current quark mass $m_0$ and constituent
quark mass $M$ via the relation
\be
M = m_0+2G_SN_cN_f\int \frac{d^3\bp}{(2\pi)^3}\frac{M}{E}\big(1-f_{Q}-f_{\bar{Q}})~,
\label{Gap_explicit}
\ee
where
\be
f_{Q,\bar{Q}}=\frac{1}{e^{(E \mp \tilde{\mu})/T}+1}~,
\label{NJL_f}
\ee
and quark number density make the quark chemical potential $\mu$
shift to an effective chemical potential
\be
\tilde\mu=\mu-G_V n~.
\label{mu_eff}
\ee 
Since, NJL is not renormalizable, we regularize the diverging vacuum integral by introducing a sharp 
three momentum cut-off $\Lambda$. The energy of the quasi-quark (both up and down) of 
constituent mass $M$ is given as $E=\sqrt{{\bp}^2+M^2}$.
The chiral condensate $\Sigma$ at finite temperature depends on Fermi-Dirac
distribution function, which is the function of effective chemical potential,
given in Eq.~(\ref{mu_eff}). Hence, $G_V$ dependence enters to the Gap equation 
through this thermodynamical phase space.
This gap Eq.~(\ref{Gap_explicit}) is plotted in Fig.~(\ref{fig:mass}) for different values of
$G_V$ and we find a mild noticeable enhancement of $M$ with $G_V$ in the intermediate
temperature range. Decreasing of quark chemical potential with $G_V$ make thermal part
shrink. Therefore, the contribution of [vacuum - thermal]-term in the left hand side of 
(self-consistent) Eq.~(\ref{Gap_explicit}) is increased, for which we are getting
a increasing trend of $M$ with $G_V$.
We can get 
back to the usual NJL Lagrangian by switching the vector interaction off.

With all these in hand, we can now write the thermodynamic potential using mean field approximation as
\bea
\Omega_{\rm{NJL}}&=&\frac{G_{S}}{2}\Sigma^2-\frac{G_{V}}{2}n^2-
2N_fN_c\int_{\Lambda}\frac{d^3\bp}{(2\pi)^3}E \nonumber\\
&-&2N_fN_cT\int\frac{d^3\bp}{(2\pi)^3} \left[{\rm{ln}}(1+e^{-(E-\tilde{\mu})/T})
\right.\nn\\
&& \left.+{\rm{ln}}(1+e^{-(E+\tilde{\mu})/T}) \right].
\eea
The thermodynamic potential 
depends on both constituent quark mass ($M$) and the effective chemical potential ($\tilde\mu$). 

\subsection{PNJL}
\label{ssec:pnjl}
So far we have considered only the chiral dynamics, by which quark to hadron phase
transition can be realized as restored to broken phases of chiral symmetry. 
Now we also incorporate the deconfinement dynamics by including Polyakov loop. 
It will give us another view, where we can see the quark to hadron phase transition
as a confinement to deconfinement phase transition.
This is formally known as PNJL 
model~\cite{Ghosh:2006qh,Mukherjee:2006hq,Ratti:2005jh,Ghosh:2007wy,Fukushima:2008wg,Ghosh:2014zra}. 
Here along with the $\Sigma$ and $n$ fields we have two more mean fields \textendash\, expectation value 
of Polyakov loop $\Phi$ and its conjugate $\bar\Phi$. $\Phi$ works as the order parameter for deconfinement 
dynamics. For two flavor the PNJL Lagrangian with vector interaction is written as
\bea
 {\mathcal{L}}_{\rm PNJL} &=& \bar{\psi}(i{\Ds}-m_0+\gamma_0\mu)\psi +
\frac{G_S}{2}[(\bar{\psi}\psi)^2
\nn\\
&+&(\bar{\psi}i\gamma_5\vec{\tau}\psi)^2]
- \frac{G_V}{2}(\bar{\psi}\gamma_{\mu}\psi)^2
\nonumber\\
&-& {\mathcal U}(\Phi[A],\bar{\Phi}[A],T),
\label{eqn:pnjl}
\eea
where where $\Ds=\gamma_{\mu} D^{\mu}$ and the covariant derivative $D^\mu=\partial^\mu-ig{\mathcal A}^\mu_a\lambda_a/2$, 
${\mathcal A}^\mu_a=\delta^{\mu 0}{\mathcal A}^a_0$ being the $SU(3)$ background fields; $\lambda_a$'s
are the Gell-Mann matrices. One should note that here only two components of the gauge field, 
corresponding to $\lambda_3$ and $\lambda_8$, will contribute. 
The effective Polyakov loop gauge potential is parameterized as
\begin{equation}
 \frac{{\mathcal U}(\Phi,\bar{\Phi},T)}{T^4} = 
    -\frac{b_2(T)}{2}\Phi\bar{\Phi} -
    \frac{b_3}{6}(\Phi^3+{\bar{\Phi}}^3) +
    \frac{b_4}{4}(\bar{\Phi}\Phi)^2,
\label{eq.potential}
\end{equation}
with
\begin{equation}
 b_2(T) = a_0 + a_1\left(\frac{T_0}{T}\right) + a_2\left(\frac{T_0}{T}\right)^2 +
    a_3\left(\frac{T_0}{T}\right)^3. \nonumber\\
\end{equation}

Values of different coefficients and parameters $a_0,\ a_1,\ a_2,\ a_3,\ b_3$ , $b_4$, $T_0$ and $\kappa$ are 
same as those given in Refs.~\cite{Ghosh:2007wy, Hansen:2006ee}. 
We should note an important point here that in the NJL model the 
color trace gives us a factor of $N_c$. In the presence of background gauge field 
the color trace is not straightforward. After some mathematical manipulation the 
color trace in PNJL model also splits out a factor of $N_c$ along with a modified 
thermal distribution function for particle and antiparticle 
which read as\cite{Hansen:2006ee,Islam:2014sea}
\bea
f_Q&=&\frac{\Phi e^{-\beta(E-\tilde \mu)}
 +2\bar{\Phi}e^{-2\beta(E-\tilde \mu)}+e^{-3\beta(E-\tilde \mu)}}
 {1+3\Phi e^{-\beta(E-\tilde \mu)}+3\bar{\Phi}e^{-2\beta(E-\tilde \mu)}
 +e^{-3\beta(E-\tilde \mu)}},
\nn\\
f_{\bar{Q}}&=&\frac{\bar\Phi e^{-\beta(E+\tilde \mu)}
 +2{\Phi}e^{-2\beta(E+\tilde \mu)}+e^{-3\beta(E+\tilde \mu)}}
 {1+3\bar\Phi e^{-\beta(E+\tilde \mu)}+3{\Phi}e^{-2\beta(E+\tilde \mu)}
 +e^{-3\beta(E+\tilde \mu)}};
 \label{PNJL_f}
\eea
respectively. We get back the usual NJL results from these distribution 
functions by putting $\Phi=\bar\Phi=1$. Thus while calculating 
different transport coefficients in the ambient of these models one needs 
to be careful. For NJL model it will be sufficient to replace 
the usual mass by the effective one. But for PNJL model one also needs to 
incorporate the modified distribution functions (See Refs.~\cite{HMPQM1}). With these modified distribution 
functions the effective mass in PNJL model reads as
\bea
M = m_0+2G_SN_cN_f\int \frac{d^3\bp}{(2\pi)^3}\frac{M}{E}\big(1-f_Q-f_{\bar{Q}}\big).
\label{Gap_explicit_pnjl}
\eea

The corresponding thermodynamic potential is written as
\begin{widetext}
 \bea
 {\Omega}_{\textrm{PNJL}} &=& {\mathcal U}(\Phi,{\bar \Phi},T) + \frac{ G_S}{2} \Sigma^2 -\frac{ G_V}{2} n^2 
 \nonumber\\
 &-&2N_fT\int \frac{d^3\bp}{(2\pi)^3} \ln \left[1+ 3\left(\Phi +{\bar \Phi}
 e^{-(E-\tilde \mu)/T} \right)e^{-(E-\tilde \mu)/T} + e^{-3(E-\tilde \mu)/T} \right ]  \nonumber \\
 &-&  2N_fT\int \frac{d^3\bp}{(2\pi)^3} \ln \left[1+ 3\left({\bar \Phi} + \Phi
 e^{-(E+\tilde \mu)/T} \right)e^{-(E+\tilde \mu)/T} + e^{-3(E+\tilde \mu)/T} \right ] \nonumber \\ 
 &-&\kappa T^4 \ln[J(\Phi,{\bar \Phi})] 
  -2N_fN_c\int_{\Lambda}\frac{d^3\bp}{(2\pi)^3}E\ .
 \label{eqn:thermo_pot_pnjl}
 \eea
\end{widetext}
The Vandermonde determinant $J(\Phi,{\bar \Phi})$ is 
given by\cite{Ghosh:2007wy,Islam:2014tea}
\begin{equation}
J[\Phi, {\bar \Phi}]=\frac{27}{24\pi^2}\left[1-6\Phi {\bar \Phi}+
4(\Phi^3+{\bar \Phi}^3)-3{(\Phi {\bar \Phi})}^2\right].
\end{equation}

\subsection{EPNJL}
\label{ssec:epnjl}
It has been confirmed through different lattice QCD simulation that chiral and deconfinement transitions take 
place at the same temperature~\cite{Fukugita:1986rr} or nearly the same temperature~\cite{Aoki:2006br}. 
Now this is not clearly understood whether it is a mere coincidence or there are some correlations between 
these two apparently distinct phenomena. To understand this coincidence through effective models a conjecture 
of strong entanglement between the chiral and deconfinement dynamics has been proposed~\cite{Sakai:2010rp,Sugano:2014pxa}.
Because of this entanglement of two dynamics it is known as EPNJL model. This is realized by introducing Polyakov 
loop dependent coupling constants, where the form of the ansatz is so chosen that it is $Z_3$ symmetric. Thus the 
Lagrangian in EPNJL model is same as that in (\ref{eqn:pnjl}) except the coupling constants $G_S$ and $G_V$ are now 
replaced by $\tilde G_S(\Phi)$ and $\tilde G_V(\Phi)$. They are given by
\begin{equation}
 \tilde{G}_S(\Phi)= G_S[1-\alpha_1\Phi\bar\Phi-\alpha_2(\Phi^3+\bar\Phi^3)]  , \label{eqn:entangle_Gs}
\end{equation}
and
\begin{equation}
 \tilde{G}_V(\Phi)= G_V[1-\alpha_1\Phi\bar\Phi-\alpha_2(\Phi^3+\bar\Phi^3)]  . \label{eqn:entangle_Gv}
\end{equation}
If we put $\alpha_1=\alpha_2=0$ we get back usual PNJL model. The strength of the vector coupling constant is, as 
mentioned earlier, taken in terms of values of $G_S$. In the same way we can get the thermodynamic potential 
for EPNJL model by introducing Polyakov loop dependent coupling constants in Eq.~(\ref{eqn:thermo_pot_pnjl}).
\begin{figure}[hbt]
\begin{center}
\includegraphics[scale=0.8]{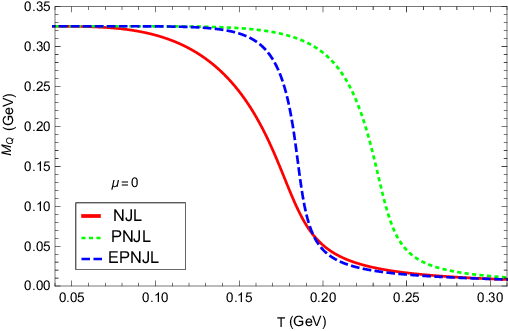}
\end{center}
\caption{Temperature dependence of constituent quark masses for 
NJL (solid line), PNJL (dotted line), EPNJL (dashed line) at $\mu=0$.}
\label{fig:mass2}
\end{figure}
Now along with all the parameters in PNJL model we have two new parameters, $\alpha_1$ and $\alpha_2$ 
which need to be fixed. This is done and discussed in details in~\cite{Islam:2015koa}. It is found there 
that the values of ($\alpha_1$,~$\alpha_2$)= (0.1, 0.1) allow to reproduce the coincidence of two transition 
temperatures to be within the range provided by lattice QCD for zero chemical 
potential~\cite{Karsch:2001cy,Karsch:2000kv}. The explicit form of the gap equation in EPNJL model 
is the same as that written in Eq.~(\ref{Gap_explicit_pnjl}) except that $G_S$ and $G_V$ will 
now be replaced by $\tilde G_S$ and $\tilde G_V$ as given in 
Eqs.~(\ref{eqn:entangle_Gs}) and~(\ref{eqn:entangle_Gv}) respectively.

The picture of transition from a current quark mass $m_0\approx 0.010$ GeV at high $T$ to 
constituent quark mass $M\approx 0.320$ GeV at low $T$ will mainly map the quark-hadron phase
transition and maximum transition of mass is occurred at transition temperature point. In different
extended NJL models, this point is shifted. Fig.~(\ref{fig:mass2}) demonstrates it nicely. 

Let us start with the discussion of transition temperature for NJL model first.
From the melting of $M(T)$ curve (red solid line) for NJL model, one can recognize roughly
the maximum melting point as $T_\Sigma\approx 0.177$ GeV (at $\mu=0$). It is only chiral
dynamics which is associated with this mass melting in NJL model, therefore, $T_\Sigma$ is popularly
known as chiral transition temperature.
As we increase $\mu$ the transition temperature keeps on decreasing. 
On the other hand in PNJL model we have both chiral and deconfinement dynamics. So essentially 
we have two phase transitions \textendash\, one is the chiral phase transition, occurred at $T_\Sigma$ 
and the other is the deconfinement phase transition, occurred at temperature $T_\Phi$ (say). In PNJL model, at $\mu=0$,
we have found $T_\Sigma =0.233$ GeV and $T_\Phi=0.228$ GeV (for $\mu=0$, $\Phi=\bar\Phi$, 
so we have $T_\Phi=T_{\bar\Phi}$) by searching the inflection points of quark condensate and Polyakov loop, respectively{\footnote{ {\it These inflection points can be found by plotting the first temperature-derivative of $\Sigma$ or $\Phi$ as a function of temperature and finding the maximum of the corresponding plot, which signifies the transition temperature $T_\Sigma$ or $T_\Phi$,  respectively. In other words, these are the points at which the curvature changes sign. The readers might look into the Refs.~\cite{Ratti:2005jh,Ghosh:2007wy,Islam:2015koa} for a detailed discussion, particularly Ref.~\cite{Islam:2015koa} which involves the same parameter set as used in the present calculation.}}}.
As we increase $\mu$ both transition temperatures decrease 
and also there is now differences between $T_\Phi$ and $T_{\bar\Phi}$ for nonzero $\mu$, though 
very small. We take average of the two temperatures ($\frac{T_\Phi+T_{\bar\Phi}}{2}$) to denote the deconfinement temperatures for nonzero values of $\mu$. Since the chiral transition temperature 
is always very close to the deconfinement transition temperature, we use the average of the 
two ($\frac{T_\Sigma+T_{\Phi}}{2}$) to denote as the critical temperature in PNJL model.
In EPNJL model with the parameter choice ($\alpha_1$,~$\alpha_2$)= (0.1, 0.1) we get $T_\Sigma=185$
MeV and $T_\Phi=183$ MeV at $\mu=0$~\cite{Islam:2015koa}.

\subsection{Thermodynamical quantities}
\label{ssec:thermo}
We see that the thermal distributions, denoted by $f_{Q,{\bar Q}}$, are taking different forms in different versions
of the model. In NJL model it is the usual FD distribution function with the effective mass ($M_Q$) and chemical potential ($\tilde\mu$), as given in Eq.~(\ref{NJL_f}). In absence of vector interaction, $\tilde\mu$ reduces to $\mu$.
When we deal with PNJL model the FD distributions transform to some modified forms, as
given in Eq.~(\ref{PNJL_f}). Apart from these palpable differences in forms, distributions in NJL and PNJL
models are also different through the constituent quark masses, which are different for these two models 
(vide Fig.~\ref{fig:mass2}). The form of the distributions remain the same in PNJL and EPNJL models, 
but quantitatively they are again different because of their differences in effective mass (Fig.~\ref{fig:mass2}).

Now, in general, if we denote $f_{Q{\bar Q}}$ as thermal distribution functions, then
we can present our different thermodynamical quantities in terms of $f_{Q{\bar Q}}$,
owing to the quasi-particle relation of statistical mechanics.
Thermodynamical quantities like pressure~$P$, the energy 
density~$\epsilon$, and net quark or baryon density~$\rho$ can be obtained from the 
quasi-particle relations~\cite{Marty}
\bea
P &=& 2N_fN_c\int \frac{d^3\bp}{(2\pi)^3}\frac{\bp^2}{3E}
\left[f_Q+f_{\bar Q}\right],
\label{P} \\
\ep &=& 2N_fN_c\int \frac{d^3\bp}{(2\pi)^3} E
\left[f_Q+f_{\bar Q}\right],
\label{epsilon} \\
\rho &=& 2N_fN_c\int \frac{d^3\bp}{(2\pi)^3}\left[f_Q - f_{\bar Q}\right].
\label{rho_Tmu}
\eea
The entropy density~$s$ and the heat function~$h$ are related to the above quantities through
the following relations:
\bea
s &=& \frac{\ep + P-\mu \rho}{T},
\label{s_eP} \\[0.3true cm]
h &=& (\ep + P)/\rho.
\label{h_eP}
\eea
Heat function $h$ is an important quantity, defined by the ratio of enthalpy density ($\epsilon + P$) to the net quark density ($\rho$). This quantity becomes divergent (unphysical) at $\mu=0$, where net quark density vanishes. However, enthalpy density $h\rho=\ep +P$ remain finite. 
\section{Transport coefficients}
\label{sec:tr}
A detail derivation of the expressions of transport coefficients from
relaxation time approximation (RTA) can be seen in 
Refs.~\cite{Chakraborty:2010fr,Hosoya:1983xm,Gavin:1985ph,Deb,Greco,Kadam_el},
and from Kubo approach in Refs.~\cite{Ghosh:2014yea,Ghosh:2016yvt,FernandezFraile:2009mi}.
In this section, we will take a revisit of RTA methodology just for a sequential description.

To calculate different transport coefficients of relativistic fluid,
the necessary macroscopic quantities are energy-momentum tensor ($T^{\mu\nu}$),
four dimensional quark/baryon current ($N^\mu$) and electric current ($J^\mu$).
Here, 4-vectors are represents by Greek letters and 3-vectors are represents by Latin letters.
If we consider that the fluid is made up of 2-flavor quark and anti-quark, then
in microscopic kinetic theory the macroscopic quantities can be expressed as
\bea
T^{\mu\nu}&=&2N_fN_c\int\frac{d^3\bp}{(2\pi)^3}\frac{p^\mu p^\nu}{E}(f_{Q} + f_{\bar{Q}})~,
\label{eq:Tmunu}
\eea
\bea
N^\mu&=&2N_fN_c\int\frac{d^3\bp}{(2\pi)^3}\frac{p^\mu}{E}(f_{Q} - f_{\bar{Q}})~ {\rm and}
\label{eq:Nmu}
\eea
\bea
J^\mu&=&2N_c \sum_{u,d}\int\frac{d^3\bp}{(2\pi)^3}\frac{p^\mu}{E}(e_Q f_{Q} + e_{\bar{Q}} f_{\bar{Q}})~,
\label{eq:Jmn} 
\eea
where flavor degeneracy $N_f=2$; color degeneracy $N_c=3$; the summation stands for 2 flavor 
quark and anti-quark to account for their charges ($e_{u,{\bar u}}=\pm 2e/3$ and $e_{d,{\bar d}}=\mp e/3$);
particle four momentum $p^\mu=(E, \bp)$; $E=\sqrt{\bp^2 + m^2}$ for particle mass $m$;
$f_{Q,\bar{Q}}$ are non-equilibrium distribution functions of quarks and anti-quarks, respectively.  
Splitting $f_{Q,\bar{Q}}$ by equilibrium (Fermi-Dirac or modified) distribution $f_{Q,\bar{Q}}^0$ and a small deviation $\delta f_{Q,\bar{Q}}$ for 
quark and anti-quark,
i.e. 
\be
f_{Q,\bar{Q}}=f_{Q,\bar{Q}}^0 + \delta f_{Q,\bar{Q}}~, 
\label{f0df}
\ee
one can separate out the ideal and dissipation part of $T^{\mu\nu}$,
$N^\mu$ and $J^\mu$ as,
\bea
T^{\mu\nu}&=&T^{\mu\nu}_0 +T^{\mu\nu}_D~,
\label{T0D}
\\
N^\mu &=& N^\mu_{0} + N^\mu_{D}~{\rm and}
\label{N0D}
\\
J^\mu &=& J^\mu_{0} + J^\mu_{D}~.
\label{J0D}
\eea

Here, reversible/ideal part of energy momentum tensor 
is $T_0^{\mu\nu} = -g^{\mu\nu} P + (\epsilon + P)u^\mu u^\nu$, and 
$N^\mu_0$, $J^\mu_0$ are that of quark/baryon charge current and electric
charge current respectively. The dissipation parts of two currents are $N^\mu_{D}$, $J^\mu_{D}$
and for energy-momentum tensor is,
\bea
T_D^{\mu\nu}&=&W^\mu u^\nu + W^\nu u^\mu \:+\: \pi^{\mu\nu} 
+\Pi^\munu ~,
\eea
where, $W^\mu$ represents energy flow, $\pi^{\mu\nu}$ and $\Pi^{\mu\nu}$
 are shear and bulk viscous stress tensor respectively. 
All the dissipative candidates $\pi^{\mu\nu}$, $\Pi^{\mu\nu}$,
$W^\mu$ and $N^\mu_{D}$ are
orthogonal to four velocity of fluid element $u^\mu$. They can be extracted
from $T^{\mu\nu}$ and $N^\mu$ by their respective connections~\cite{Muronga}:
\bea
\pi^{\mu\nu}&=&\{\frac{1}{2}(\Delta^\mu_\sigma\Delta^\nu_\rho +\Delta^\mu_\rho\Delta^\nu_\sigma)
-\frac{1}{3}\Delta^{\mu\nu}\Delta_{\sigma\rho} \}T^{\sigma\rho}~,
\nn\\
\Pi + P &=& -\frac{1}{3}\Delta_{\mu\nu}T^\munu~,
\nn\\
q^\mu &=& W^\mu - h N_D^\mu = u_\nu T^{\nu\sigma}\Delta^\mu_\sigma
- h \Delta^\mu_\nu N^\nu~.
\label{pq_TN}
\eea
Here, projection operator orthogonal to fluid velocity is $\Delta^{\mu\nu}=g^{\mu\nu}-u^\mu u^\nu$, $\Pi$ and $P$ are  respectively
bulk and local isotropic pressure. 
In practice, four-velocity $u^\mu$ is chosen in two ways, known as Eckart and Landau frames.
In Eckart frame $u^\mu$ is parallel to $N^\mu$ and so, $N^\mu_D=0$. Similarly,
$W^\mu=0$ in Landau frame. For a system with no net charge, the four-velocity in the
Eckart formalism is not well defined. Hence, in general under this situation one should 
use the Landau frame.

The transport coefficients $\eta, \zeta, \kappa$ and $\sigma$ are basically proportionality constants, 
which make connection between thermodynamical forces ($\pi^{\mu\nu}$,
$\Pi^{\munu}$, $q^\mu$, $E^\mu$) and the corresponding 
currents ($\cal U^{\mu\nu}_\eta$, $\cal U^{\munu}_\zeta$, $\cal U^\mu_\kappa$, $J^\mu_D$) 
as,~\cite{Chakraborty:2010fr,Hosoya:1983xm,Gavin:1985ph,FernandezFraile:2009mi,Deb,Greco,Kadam_el,Greiner}
\bea
\pi^{\mu\nu}&=&\eta {\cal U}^{\mu\nu}_\eta~ ,
\nn\\
{\rm with}~{\cal U}^{\mu\nu}_\eta&=&
\FB{D^\mu u^\nu + D^\nu u^\mu -\frac{2}{3}\Delta^{\mu\nu}\partial_\rho u^\rho}~;
\label{ESeta}
\\
\Pi^\munu =\Pi \Delta^\munu&=& \zeta {\cal U}^{\mu\nu}_\zeta ,
\nn\\
{\rm with}~{\cal U}^{\mu\nu}_\zeta&=&
\Delta^{\mu\nu}\partial_\rho u^\rho~{\rm and} ~
\Pi = \zeta \partial_\rho u^\rho
\label{ZSeta}
\\
q^\mu&=&\kappa {\cal U}^{\mu}_\kappa~,
\nn\\
{\rm with}~{\cal U}^{\mu}_\kappa &=&T \Delta^{\mu\nu}\Big(\frac{D_\nu T}{T}-\frac{D_\nu P}{hn}\Big)~;
\label{ESkappa}
\\
J_D^\mu&=&\sigma^{\mu\nu} E_\nu~,~{\rm with}~E^\nu=(0,E^i)~.
\label{ESel}
\eea
Here, $D^\mu=\partial^\mu - u^\mu u^\sigma \partial_\sigma$ and,
 $E^\mu\equiv F^{0\mu}$ contains electric field part only of electromagnetic field tensor $F^{\mu\nu}$. Using Gibbs-Duhem relation,
\be
\frac{D_\nu P}{\rho}=h\frac{D_\nu T}{T}- TD_\nu \Big(\frac{\mu}{T}\Big)~,
\label{GD}
\ee
Eq.~(\ref{ESkappa}) can be further simplified as
\be
{\cal U}^{\mu}_\kappa =\frac{T^2}{h} \Delta^{\mu\nu}D_\nu \Big(\frac{\mu}{T}\Big)~.
\ee

Now, owing to the microscopic relations, given in Eq.~(\ref{eq:Tmunu}), Eq.~(\ref{eq:Nmu}) and Eq.~(\ref{eq:Jmn}), 
and then using Eq.~(\ref{pq_TN}),
we can get
\bea
\pi^{\mu\nu}&=&2 N_f N_c \Big[\frac{1}{2}(\Delta^\mu_\sigma\Delta^\nu_\rho +\Delta^\mu_\rho\Delta^\nu_\sigma)
\nn\\
&&-\frac{1}{3}\Delta^{\mu\nu} \Delta_{\sigma\rho}\Big]
\int{\frac{d^3\bp}{(2\pi)^3}\frac{p^\sigma p^\rho}{E}(\delta f_{Q} + \delta f_{\bar{Q}})}
\label{pi_df}
\\
\Pi&=& {2 N_f N_c} \Big[-\frac{1}{3} \Delta_\munu \Big]
\int{\frac{d^3\bp}{(2\pi)^3}\frac{p^\mu p^\nu}{E}} 
(\delta f_{Q} + \delta f_{\bar{Q}}) \nn\\
\label{Pi_df}
\\
q^{\mu}&=&2N_f N_c \int \frac{d^3\bp}{(2\pi)^3} \Delta^\mu_\sigma\Big[\{u_\nu\frac{p^\nu p^\sigma}{E}
-h\frac{p^\sigma}{E}\}\delta f_{Q}
\nonumber\\
&&~~~~~~~~~  +\{u_\nu\frac{p^\nu p^\sigma}{E}
+h\frac{p^\sigma}{E}\}\delta f_{\bar{Q}}\Big]
\label{I_df}
\\
{\rm and}~J_D^\mu&=&2N_c \sum_{Q=u,d}\int\frac{d^3\bp}{(2\pi)^3}\frac{p^\mu}{E} \:
(e_Q \delta f_Q + e_{\bar{Q}} \delta f_{\bar{Q}})~.
\label{J_df}
\eea

In local rest frame, four velocity $u = (1, {\boldsymbol 0})$, $p.u=E$,
and hence, Eq.~(\ref{I_df}) can be written as
\bea
q^{\mu}&=&2N_f N_c \int \frac{d^3\bp}{(2\pi)^3} \frac{p^\mu}{E}
\nonumber\\
&& \Big\{(E- h) \delta f_{Q} +(E + h) \delta f_{\bar{Q}}\Big\}~.
\label{I_df2}
\eea


The small deviation of the (Fermi-Dirac or modified) distribution function can be assumed as 
\bea
\delta f_{Q,\bar{Q}}
=(f_{Q,\bar{Q}}-f_{Q,\bar{Q}}^0) &\propto& -\frac{\del f_{Q,\bar{Q}}^0}{\del E}
\nn\\
&\propto&\beta f_{Q,\bar{Q}}^0(1-f_{Q,\bar{Q}}^0)
\nn\\
&=&\phi^{\FB{Q,\bar{Q}}} \: \beta f_{Q,\bar{Q}}^0(1-f_{Q,\bar{Q}}^0)~,
\nn\\
\label{df_phi}
\eea
where $\phi^{\FB{Q,\bar{Q}}}$ will contribute to dissipative part
of energy-momentum tensor $T^{\mu\nu}_D$, quark/baryon charge current $N^\mu_D$,
electric charge current $J^\mu_D$, as defined in Eqs~(\ref{f0df})-(\ref{J0D}).
To satisfy Landau-Lifshitz condition $u^\mu T^{\mu\nu}_D=0$, a natural choice
is to use the same tensorial decomposition, as defined in Eqs.~(\ref{ESeta})-(\ref{ESel}).
Hence $\phi^{\FB{Q,\bar{Q}}}$ can be expressed as a function of space time
and momentum as~\cite{Chakraborty:2010fr,Gavin:1985ph,Deb,Greco,Kadam_el,Greiner}
\bea
\phi^{\FB{Q,\bar{Q}}}&=&A_{\mu\nu}^{\FB{Q,\bar{Q}}} {\cal U}^{\mu\nu}_\eta 
+ B_{\mu}^{\FB{Q,\bar{Q}}} {\cal U}^{\mu}_\kappa 
+C_{\mu}^{\FB{Q,\bar{Q}}} {E}^{\mu} \nn\\
&& +Z^{\FB{Q,\bar{Q}}}(\partial_\rho u^\rho)  ~.
\label{phi_A}
\eea

The coefficient factors $A_{\mu\nu}$, $B_{\mu}$, $C_{\mu}$
and $Z$ for different thermodynamical tensors ${\cal U}^{\mu\nu}_\eta$,
${\cal U}^{\mu}_\kappa$, $E^{\mu}$ and $(\partial_\rho u^\rho)$ are associated with corresponding 
transport coefficients $\eta$, $\kappa$, $\sigma$ and $\zeta$ respectively.
These coefficient factors can be obtained with the help of Boltzmann equation,
\bea
\frac{\del f_{Q,\bar{Q}}}{\del t}+\frac{\del x^i}{\del t}\frac{\del f_{Q,\bar{Q}}}{\del x^i}+\frac{\del p^i}{\del t}\frac{\del f_{Q,\bar{Q}}}{\del p^i}
&=&\Big(\frac{\del f_{Q,\bar{Q}}}{\del t}\Big)_{\rm col}
\nn\\
\Rightarrow\frac{\del f_{Q,\bar{Q}}}{\del t}+v^i\frac{\del f_{Q,\bar{Q}}}{\del x^i}+F^i\frac{\del f_{Q,\bar{Q}}}{\del p^i}
&=&\Big(\frac{\del f_{Q,\bar{Q}}}{\del t}\Big)_{\rm col}~,
\nn\\
\label{BE}
\eea
where $v^i$ is particle velocity, $F^i$ is applied force on the particle.
The right hand side of Eq.~(\ref{BE}), representing the collisional term (accounting 
for the forces acting between particles in collisions), can be approximated by
Anderson-Witting collision term~\cite{Anderson:1973ph},
\be
\Big(\frac{\del f_{Q,\bar{Q}}}{\del t}\Big)_{\rm col}=-\Big(\frac{p^\mu u_\mu}{E}\Big)
\frac{\delta f_{Q,\bar{Q}}}{\tau_{Q, \bar{Q}}}~.
\label{RTA_df}
\ee
This is standard relaxation time approximation (RTA) technique, where $\tau_{Q, \bar{Q}}$ is the rough time scale
required for the particle/anti-particle to relax from its non-equilibrium distribution $f_{Q,\bar{Q}}$
to equilibrium distribution $f^0_{Q,\bar{Q}}$. 
Using Eq.~(\ref{RTA_df}) in Eq.~(\ref{BE}) and then express that RTA based Boltzmann transport equation in covariant form as
\be
\frac{1}{E}p^\mu \del_{\mu} f_{Q,\bar{Q}}+e_{Q,\bar{Q}}F^{\munu}\frac{p_\nu}{E}\frac{\del f_{Q,\bar{Q}}}{\del p^\mu}
=-\Big(\frac{p^\mu u_\mu}{E}\Big)\frac{\delta f_{Q,\bar{Q}}}{\tau_{Q, \bar{Q}}}~.
\label{RBE}
\ee
In right hand side (rhs), $\delta f_{Q,\bar{Q}}$ can be expressed in terms of 
corresponding tensors (${\cal U}^{\mu\nu}_\eta$, $(\partial_\rho u^\rho)$, ${\cal U}^{\mu}_\kappa$ and $E^{\mu}\equiv F^{0\mu}$), 
associated with the transport coefficients ($\eta$,
$\zeta$, $\kappa$ and $\sigma$), by using the Eq.~(\ref{phi_A}). 
In left hand side (lhs), we will assume $f_{Q,\bar{Q}}\approx f^0_{Q,\bar{Q}}$.
Let us proceed for FD distribution of NJL model but the same steps can be done for modified distribution of PNJL/EPNJL model if we follow the Appendix, given in Sec.~(\ref{sec:App2}). We can write FD distribution in covariant form:
\be
f^0_{Q,\bar{Q}}=1/\{{\rm exp}\Big(\frac{p^\mu u_\mu\mp\mu}{T}\Big)+1\}~,
\label{f_FD}
\ee
where $p^\mu$ is particle quantity (four momentum) and $u^\mu$, $T$, $\mu$ are fluid
quantities, which depend on space and time. So Eq.~(\ref{RBE}) will get the modified form:
\bea
\frac{1}{E}p^\mu \del_{\mu} f^0_{Q,\bar{Q}} &+& e_{Q,\bar{Q}}F^{\munu}\frac{p_\nu}{E}\frac{\del f^0_{Q,\bar{Q}}}{\del p^\mu}
=-\Big(\frac{p^\mu u_\mu}{E T}\Big)\frac{1}{\tau_{Q, \bar{Q}}}
\nn\\
&&\Big[A_{\mu\nu}^{\FB{Q,\bar{Q}}} {\cal U}^{\mu\nu}_\eta
 + B_{\mu}^{\FB{Q,\bar{Q}}} {\cal U}^{\mu}_\kappa 
+C_{\mu}^{\FB{Q,\bar{Q}}} {E}^{\mu}
 \nn\\
 &&~~+ Z^{\FB{Q,\bar{Q}}} (\partial_\rho u^\rho) \Big]
f^0_{Q,\bar{Q}}(1-f^0_{Q,\bar{Q}}).~
 \label{RBE2}
\eea
Now, the idea is to express the lhs of Eq.~(\ref{RBE2}) in terms of the tensors, sitting in rhs,
so that we can equate their coefficients in both side and get the expressions of $A^{\mu\nu}$, $B^\mu$, $C^\mu$ and $Z$.
The first term of lhs in Eq.~(\ref{RBE2}) can be expressed in terms of ${\cal U}^{\mu\nu}_\eta$, ${\cal U}^{\mu}_\kappa$ and $\partial_\rho u^\rho$~\cite{Gavin:1985ph,Chakraborty:2010fr,Deb}, 
whereas second term of lhs in Eq.~(\ref{RBE2}) can be expressed in terms of $E^{\mu}\equiv F^{0\mu}$~\cite{Greco,Kadam_el,Greiner},
and then one can find,
\bea
A_{\mu\nu}^{\FB{Q, \bar{Q}}}&=&\tau_{Q,\bar{Q}}\frac{ \: p_\mu p_\nu}{E} ~,
\nn\\
B_{\mu}^{\FB{Q,\bar{Q}}}&=&\tau_{Q,\bar{Q}}\frac{\beta \: p_\mu}{E}\Big(E \mp h \Big),
\nn\\
C_{\mu}^{\FB{Q,\bar{Q}}}&=&\tau_{Q,\bar{Q}}\frac{e_{Q, {\bar Q}} \: p_\mu}{E}~
\nn\\
\rm{and,}~ 
Z^{\FB{Q, \bar{Q}}}&=&\tau_{Q,\bar{Q}}
\frac{1}{3E} \left[\bp^2- 3c_s^2\left(E^2-T^2\frac{dM_Q^2}{dT^2}\right)\right]~, \nn\\
\label{A_tr}
\eea 
where bulk viscosity component $Z^{\FB{Q,\bar{Q}}}$ is obtained for $\mu=0$, but components of shear viscosity ($A_{\mu\nu}^{\FB{Q, \bar{Q}}}$) and electrical conductivity ($C_{\mu}^{\FB{Q,\bar{Q}}}$) can be used for both $\mu=0$ and $\mu\neq 0$ (just by changing distribution function). The component of thermal conductivity $B_{\mu}^{\FB{Q,\bar{Q}}}$ is relevant for $\mu\neq 0$ as it carries the quantity - enthaply per particle $h$, which is diverged at $\mu=0$.  
The detail calculation of above outcome is given in Appendix~(\ref{sec:App1}).

Now, using Eqs.~(\ref{A_tr}) in Eq.~(\ref{phi_A}), Eq.~(\ref{df_phi}) and then in
Eqs.(\ref{pi_df}, \ref{Pi_df}, \ref{J_df}, \ref{I_df2}), we get 
\begin{widetext}
\bea
\pi^{\mu\nu} &=& 2N_FN_c\{\frac{1}{2}(\Delta^\mu_\sigma\Delta^\nu_\rho +\Delta^\mu_\rho\Delta^\nu_\sigma)
-\frac{1}{3}\Delta^{\mu\nu}_{\sigma\rho}\}
\int \frac{d^3\bp}{(2\pi)^3}\Big(\frac{p^\sigma p^\rho}{E}\Big)\beta\Big(\frac{p^\alpha p^\beta}{E}\Big)
\Big\{\tau_{Q}{f^0_Q}  (1-f^0_Q)
+ \tau_{\bar{Q}}f^0_{\bar Q} (1-f^0_{\bar Q})\Big\}{\cal U}_{\alpha\beta}^\eta~,
\nn\\
\label{pi_t} 
\\
\Pi &=& 2N_FN_c
\int \frac{d^3\bp}{(2\pi)^3}
\frac{\beta}{9E^2} \left[\bp^2- 3c_s^2\left(E^2-T^2\frac{dM_Q^2}{dT^2}\right)\right]^2
\Big\{\tau_{Q}{f^0_Q}  (1-f^0_Q)
+ \tau_{\bar{Q}}f^0_{\bar Q} (1-f^0_{\bar Q})\Big\}(\partial_\rho u^\rho)~,
\label{Pi} 
\\
q^\mu &=& 2N_FN_c\int \frac{d^3\bp}{(2\pi)^3}
\left(\frac{p^\mu}{E}\right)\beta^2\left(\frac{p^\nu}{E}\right)
\Big\{\tau_{Q} (E-h)^2 f^0_Q (1-f^0_Q) + \tau_{\bar{Q}}(E+h)^2f^0_{\bar Q}
(1-f^0_{\bar Q})\Big\}{\cal U}_{\nu}^\kappa~,
\label{q_t}
\\
{\rm and} \: \:J^\mu &=& 2N_c\sum_{Q=u,d}
\int \frac{d^3\bp}{(2\pi)^3} \left(\frac{p^\mu}{E}\right)\beta\left(\frac{p^\nu}{E}\right) 
\Big\{e_Q^2\tau_{Q}f^0_Q 
(1 -f^0_Q)
+ e_{\bar{Q}}^2\tau_{\bar{Q}}f^0_{\bar Q} (1-f^0_{\bar Q})\Big\}E_\nu ~.
\label{J_t}
\eea
\end{widetext}
Now, if we compare Eqs.(\ref{pi_t}), (\ref{Pi}), (\ref{q_t}) and (\ref{J_t}) with Eqs.~(\ref{ESeta}), (\ref{ZSeta}),
(\ref{ESkappa}) and (\ref{ESel}), then 
we can identify the final expressions of
transport coefficients as
\begin{widetext}
	\bea
	\eta &=& \frac{2N_FN_c\beta}{15}
	\int \frac{d^3\bp}{(2\pi)^3}\left(\frac{\bp^2}{E}\right)^2
	\Big\{\tau_{Q}{f_Q}  (1-f_Q)
	+ \tau_{\bar{Q}}f_{\bar Q} (1-f_{\bar Q})
	\Big\} ,
	\label{eta} 
	\\
	\zeta &=& \frac{2N_FN_c\beta}{9}
	\int \frac{d^3\bp}{(2\pi)^3}\frac{1}{E^2} \left[
	\bp^2-3c_s^2 \left(E^2-M_Q T\frac{dM_Q}{dT}\right) \right]^2
	\Big\{\tau_{Q}{f_Q}  (1-f_Q)
	+ \tau_{\bar{Q}}f_{\bar Q} (1-f_{\bar Q})
	\Big\} ,
	\label{zeta} 
	\\
	\kappa &=& \frac{2N_FN_c\beta^2}{3}\int \frac{d^3\bp}{(2\pi)^3}
	\left(\frac{\bp}{E}\right)^2
	\Big\{\tau_{Q} (E-h)^2 f_Q (1-f_Q) + \tau_{\bar{Q}}(E+h)^2f_{\bar Q}
	(1-f_{\bar Q})
	\Big\} 
	\label{kappa}
	\\
	{\rm and,} \: \:\sigma &=& \left(\frac{2N_c\beta}{3}\right) \left(\frac{5 e^2}{9}\right)
	\int \frac{d^3\bp}{(2\pi)^3} \left(\frac{\bp}{E}\right)^2 
	\Big\{\tau_{Q}f_Q 
	(1 -f_Q)
	+ \tau_{\bar{Q}}f_{\bar Q} (1-f_{\bar Q})\Big\} ~.
	\label{sigma}
	\eea
\end{widetext}
Here, speed of sound, $c_s^2=\frac{s}{T\left(\frac{ds}{dT}\right)_V}$.
To simplify the notation, we have put $f_{Q, {\bar Q}}$ in last expressions instead of $f^0_{Q, {\bar Q}}$.
We will consider $f_{Q, {\bar Q}}$ as equilibrium distribution function for the last expressions 
and also all other sections and subsections. FD distribution will be taken as equilibrium distribution
for NJL model, while modified distribution, given in Eq.~(\ref{PNJL_f}), will be considered
as equilibrium distribution for PNJL or EPNJL models. This replacement calculation for transport
coefficients can be found in Ref.~\cite{HMPQM1}.
Here also we have addressed the same in Appendix~(\ref{sec:App2}).

\subsection{Causal Aspects}
\label{sec:causal}
Now, above diffusion relations, Eq.~(\ref{ESeta}), (\ref{ESkappa}), (\ref{ESel}) don't carry any
time information, they are instantaneous relations and therefore violate the causality. 
Among the huge number of references on it, readers can follow Ref.~\cite{Romatchke:2009,Denicol,IS79,th_causal,eta_s_NEWS,Muronga,
	el_causal} for causal aspects in viscosity, thermal conductivity and electrical
conductivity. Here, we will go through causal aspects in shear viscosity estimation only.

To understand acausality problem of Navier-Stokes equation, 
if we can consider a very small perturbation in energy density $\epsilon\rightarrow\epsilon +\delta\epsilon$ and 
fluid velocity $u^{\mu}\rightarrow u^\mu +\delta u^\mu$, 
then we will get a dispersion relation of diffusion Eq.~(\ref{ESeta}) as ~\cite{Romatchke:2009},
\be
\om = \frac{\eta}{\epsilon + p}k^2~,
\ee
where $k$ is wave vector.
Hence, we can get diffusion speed,
\be
v_T(k) = \frac{d\om}{dk} = 2\frac{\eta}{\epsilon + P}k ~,
\ee
which means diffusion speed can be infinite (by crossing speed of light) as $k$ tends to infinite.
The Navier-Stokes equation (\ref{ESeta}) is actually derived from 1st order thermodynamics.
But if we consider entropy density up to second order then we can obtain causal hydrodynamics equations
[\cite{IS79,W.Hiscock,Romatchke:2009}]
:
\bea
\nn \pi^{\munu} = \eta (D^\mu u^\nu + D^\nu u^\mu - \frac{2}{3}\Delta^{\munu} D_\alpha u^\alpha + \pi^{\munu} TD(\beta_2/T) \\
- 2 \beta_2 D \pi^{\munu} - \beta_2 \pi^{\munu} \partial_{\alpha} u^{\alpha})~~~.
\label{IS}
\eea

Which is causal replacement of eq.~(\ref{ESeta}).
Realizing the new coefficient $\beta_2$ as, $\beta_2=\frac{\tau_\pi}{2\eta}$ where, $\tau_{\pi}$ is defined as shear relaxation time, we can get dispersion relation for eq.~(\ref{IS}) as ~\cite{Romatchke:2009},
\be
\om - i\frac{\eta}{\epsilon + p} \frac{k^2}{1+i\om \tau_{\pi}} =0~.
\ee
%
%
Then the diffusion speed at very large $k$  becomes:
\be
v_T^{max} \equiv \lim\limits_{k\rightarrow \infty} \sqrt{\frac{\eta}{(\epsilon + P)\tau_{\pi}}}~.
\label{vmax_tpi}
\ee
%
Here, the subscript $T$ stands for transverse velocity. The diffusion speed will not be greater than the speed of light if 
\be
\tau_{\pi}\geq \frac{\eta}{\epsilon + P}~,
\label{tpi_eta}
\ee
which is observed for all known fluid. One can recover the instantaneous Eq.~(\ref{ESeta}) by using
$\tau_{\pi}=0$. This fact will be well explored in Sec.~(\ref{sec:causal}) with different extensions of NJL model.

\section{Results}
\label{sec:Res}
\subsection{$\mu=0$ case}
\begin{figure} [ht]
\includegraphics[scale=0.8]{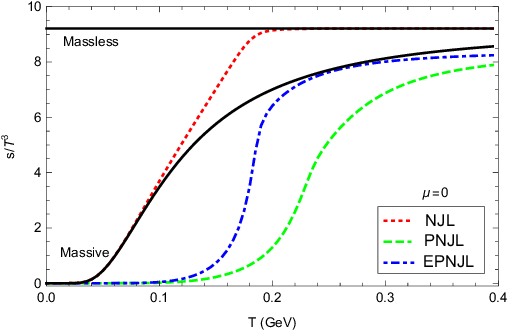}
\caption{Normalized entropy density $s/T^3$ vs temperature $T$ for massive quarks based on NJL (dotted line), EPNJL (dash-dotted line) and PNJL (dashed line) models. As limiting cases we have also shown plots for massless (horizontal solid black line) and massive (black solid line) quarks using standard F-D distribution function. For massive quarks we used confinement mass, which is calculated to be $326$ MeV.}
\label{fig:sT3}
\end{figure}

In Sec.~\ref{sec:model}, we have discussed about the formalism of different
extension components of NJL models like (a) vector interaction (\ref{ssec:njl}), 
(b) PNJL (\ref{ssec:pnjl}) and (c) EPNJL (\ref{ssec:epnjl}). Present article is
intended to investigate the comparative role of these different extensions of NJL models
on transport coefficients of quark matter, where we will discuss about the results 
for $\mu=0$ case in this sub-section. Before that, let us see the thermodynamical
quantity like entropy density, which will be required to measure the fluid property of quark matter.
The governing expression is Eq.~(\ref{s_eP}).
Fig.~(\ref{fig:sT3}) shows the $T$ dependence of (normalized) entropy density, where a straight 
horizontal line (black solid line) denotes its massless value ($s\approx 9.2 T^3$), commonly known as
Stefan-Boltzmann (SB) limit. Now, the interaction reduces that value as shown by
dotted red, dash-dotted blue and dashed green lines in Fig.~(\ref{fig:sT3}), which are obtained from 
NJL, EPNJL and PNJL models, respectively. Through these different
extended effective QCD models, interaction is mainly mapped through $T$ dependent masses $M(T)$, shown
in earlier Fig.~(\ref{fig:mass2}). Since the thermodynamical phase space part of $s$ is mainly controlled by
$M(T)$, so one can mark a similar kind of transition pattern between $M(T)$ and $s(T)$.  
High $T$ entropy density $s$ of PNJL and EPNJL models are suppressed from SB limits because the FD distributions
of NJL model are replaced by their respective modified Polyakov loop distribution. We have also added a curve for constant confinement mass $M=0.326$ GeV, shown by black solid line, drawn using standard Fermi-Dirac distribution function. It is included to demonstrate that the effective models recover this limiting value at small temperature. 
Thus all the models (NJL, PNJL and EPNJL) basically reproduce or approach the massive and massless limits at $T\rightarrow 0$ and $T\rightarrow\infty$, respectively.      

\begin{figure} [ht]
\includegraphics[scale=0.8]{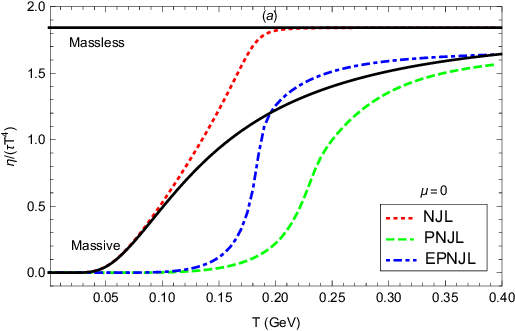}
\includegraphics[scale=0.8]{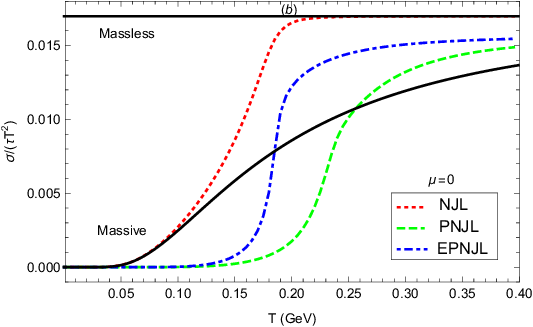}
\includegraphics[scale=0.28]{bulk.eps}
\caption{Normalized (a) shear viscosity $\eta/(\tau T^4)$, (b) electrical conductivity 
$\sigma/(\tau T^2)$, (c) bulk viscosity $\zeta/(\tau T^4)$ vs temperature $T$ for massive quarks based on NJL (dotted line), EPNJL (dash-dotted line) and PNJL (dashed line) models. As limiting cases we have also shown plots for massless (horizontal solid black line) and massive (black solid line) quarks using standard F-D distribution function. For massive quarks we used confinement mass, which is calculated to be $326$ MeV.}
\label{fig:ShT4}
\end{figure}

Next, we come to the transport coefficients estimations from Eqs.~(\ref{eta}), (\ref{kappa}), (\ref{sigma})
and (\ref{zeta}).
If we notice the expressions of transport
coefficients in Eqs~(\ref{eta}), (\ref{sigma}), (\ref{kappa}), (\ref{zeta}) then we can identify
two parts, carrying temperature ($T$) and chemical potential ($\mu$) dependent 
information. One is relaxation time of medium constituent and another is the 
thermodynamical part, influenced by its Fermi-Dirac/modified distribution function as well as
$T$, $\mu$ dependent mass. At $\mu=0$, shear viscosity $\eta$, 
bulk viscosity $\zeta$ and electrical conductivity $\sigma$
are relevant transport coefficients, as thermal conductivity $\kappa$ is diverged/not well-defined at $\mu=0$.
Hence, to zoom in the thermodynamical phase-space part of $\eta$, $\sigma$ and $\zeta$, we have plotted
$\eta/(\tau T^4)$, $\sigma/(\tau T^2)$ and $\zeta/(\tau T^4)$ vs $T$ in Fig.~\ref{fig:ShT4}(a), (b) and (c) respectively.
Interestingly, we can find similar kind of pattern in $\eta/(\tau T^4)$, $\sigma/(\tau T^2)$ as
we have found for $s/T^3$. It is because all are basically mapping approximately
similar kind of (normalized) thermodynamical phase-space components. Therefore, according to
their rapid changing point in temperature axis, different extended models follow same ranking.
NJL melts first at low $T$, then EPNJL and then PNJL at relatively high $T$.
Bulk viscosity in Fig. {\ref{fig:ShT4}}(c) shows peaks near the
transition temperatures of the respective models, as expected~\cite{G_IFT}. There will be two sources,
for which bulk viscosity contribution becomes maximum near transition temperature.
First and dominant sources is the interaction measure of thermodynamics $\ep -3P$, which
is vanished in massless medium or high temperature QCD medium but become non-zero in the intermediate and 
low temperature regions. LQCD as well as effective QCD model calculations like NJL exhibit maximum interaction
measure near transition temperature. Being proportional with interaction measure ($\zeta\propto(\ep -3P)$), 
bulk viscosity displays similar kind of peak structure near transition temperature. This interaction measure
can alternatively be understood as $(\frac{1}{3}-c_s^2)$, which basically interprets the deviation of
speed of sound square from its massless value $1/3$. This $\zeta\propto(\frac{1}{3}-c_s^2)\propto(\ep -3P)$ relation is thus main source for exhibiting the peak pattern of bulk viscosity, which alternatively reveals the conformal breaking structure of QCD medium~\cite{G_IFT}. Another source is the quantity $\frac{dM}{dT}$, which shows peak structure near chiral transition temperature $T_{\Sigma}$. Sitting in the expression of bulk viscosity, $\frac{dM}{dT}$ and $(\frac{1}{3}-c_s^2)$ become two sources to amplify the peak structure.    
\begin{table}[htb]
\caption{Locations of temperatures (MeV), where different temperature dependent quantities -
quark condensate $\Sigma$, Polyakov loop $\Phi$, $s/T^3$, $\eta/(\tau T^4)$, $\sigma/(\tau T^2)$ show
their change/transition (for $\mu=0$ case).}
\begin{ruledtabular}
\begin{tabular}{l|ccccc}
Models & $\Sigma$ &  $\Phi$ & $s/T^3$ & $\eta/(\tau T^4)$ & $\sigma/(\tau T^2)$ \\
\hline
NJL & 177 & - & {\rm two~peak} & 165 & 172  \\
PNJL & 233 & 228 & 230 & 230 & 232  \\
EPNJL & 185 & 183 & 184 & 183 & 185  \\
\end{tabular}
\end{ruledtabular}
\label{tab1}
\end{table}

At the end of Sec.~(\ref{ssec:epnjl}), we have discussed about quark condensate $\Sigma$
{\footnote{not to confused with electrical conductivity}}
and Polyakov loop $\Phi$, which change with $T$ to map the chiral and confinement-deconfinement
phase transitions, respectively. The order parameter ($\Sigma$) can be estimated from NJL model, while
PNJL and EPNJL can describe both order parameters ($\Sigma$, $\Phi$). The transition temperatures
$T_\Sigma$ and $T_\Phi$ are basically the inflection points, which are calculated by taking the derivative of the corresponding order parameter with respect to $T$ and then finding the extremum for that. The quantities $s/T^3$, $\eta/(\tau T^4)$ and $\sigma/(\tau T^2)$
are quite interesting as they contain collective effect of both the order parameters. Table~(\ref{tab1}) documents the values of these temperatures for order parameters - quark condensate $\Sigma$, Polyakov loop $\Phi$ and as well as for
the quantities $s/T^3$, $\eta/(\tau T^4)$ and $\sigma/(\tau T^2)$. The temperatures for the transport coefficients and the entropy density have been estimated in the same fashion as it is done for the order parameters. From the expressions of $s/T^3$, $\eta/(\tau T^4)$ and $\sigma/(\tau T^2)$,
written above, one notices that $\Sigma(T)$ enters through $M(T)$, while $\Phi(T)$ enters through both
$M(T)$ and thermal distribution functions.
Though two order parameters enter in the expressions of $s/T^3$, $\eta/(\tau T^4)$ and $\sigma/(\tau T^2)$ in the same ways, but 
their momentum dependent integrands are different and therefore, they are not showing the 
same temperatures, as evident from table~(\ref{tab1}). The differences are more evident for NJL model; as one introduces the background gauge field in PNJL or EPNJL model the differences almost vanish and the temperatures calculated from these quantities are almost similar to those calculated from the order parameters. Only $s/T^3$ in NJL model exhibit two peak structure 
instead of one peak, which is a model-parameter dependent fact. So, ignoring this fact
we can roughly conclude that the transition points of $s/T^3$, $\eta/(\tau T^4)$ and $\sigma/(\tau T^2)$ 
are close to average values of
$T_\Sigma$ and $T_\Phi$ for PNJL and EPNJL models.

\subsection{Perfect fluid and Causal aspects}
\label{sec:causal}
\begin{figure} [ht]
\includegraphics[scale=0.8]{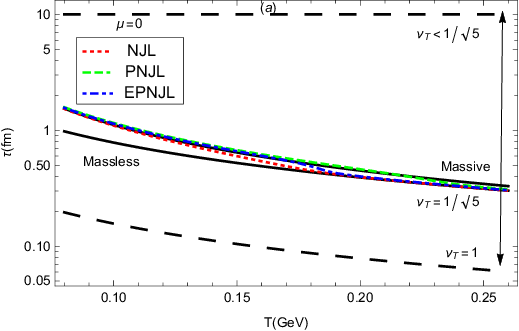}
\includegraphics[scale=0.8]{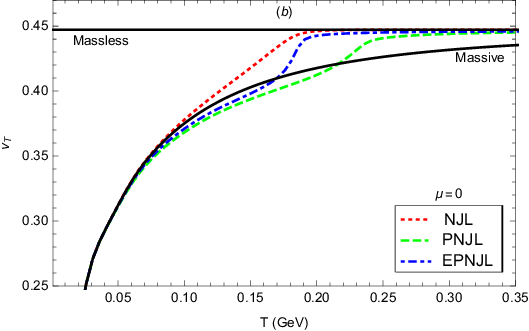}
\caption{(a)Temperature dependence of relaxation time in NJL, PNJL and EPNJL models, for which we get $\eta/s=1/4\pi$. For comparison we have also given plots for massless and massive (with confinement mass, $326$ MeV) quarks, using standard F-D statistics. (b) Corresponding maximum values of diffusion speed $v_T$.}
\label{fig:Tau_KSS}
\end{figure}

We have normalized information of $\tau$ during plotting shear viscosity in Fig.~(\ref{fig:ShT4}), 
but it can also be a temperature dependent quantity, if one attempts to calculate it microscopically
and $\tau(T)$ can modify the $T$ dependent profile of shear viscosity as well as other transport coefficients.
From experimental side, $\eta/s$ of quark matter created at RHIC is found to be very
close to its lower bound $\frac{1}{4\pi}$, based on viscous hydrodynamic 
model analysis of elliptic flow~\cite{hydro_rev}.
We may get a rough idea about the values of $\tau$, for which our estimated
$\eta/s$ will be close to the lower bound. This restriction also give us a
temperature dependent $\tau$ instead of its constant value.
%
%
For massless spin 1/2 particle with zero chemical potential, $\tau=5/4\pi T$ give us $\eta/s=1/4\pi$.
This is shown as the black line in the Fig.~\ref{fig:Tau_KSS}. 
Imposing same restriction of $\eta/s=1/4\pi$ in NJL, PNJL and EPNJL model calculations, 
we get required relaxation time $\tau(T)$, displayed by dotted, dashed and dash-dotted
lines in Fig.~(\ref{fig:Tau_KSS}).
Let us analyze these curves. 
We know that (approximately) massless quark can only be expected at very high temperature but 
as we decrease the temperature, the non-zero quark condensate will form, for 
which constituent quark mass also grows up. Mapping this fact via gap equation in NJL model,
thermodynamical part of $\eta$ become suppressed in low temperature domain with respect to the massless case. This lower value of thermodynamical part can be compensated by little higher
values of $\tau$ for getting same values of $\eta/s$ ($=1/4\pi$) as obtained in massless case.
Therefore red dotted line ($\tau$ of NJL model) is quite larger than black solid line
($\tau$ of massless case) in low temperature domain. Above the transition temperature,
both curves are merged as condensate melts down completely.
When we transit to PNJL model, the confinement picture has been taken into consideration (statistically) via modified
thermal distribution function, which has lower statistical weight than FD distribution.
So, with respect to NJL case, PNJL has lower strength for thermodynamical part of $\eta$, so for getting KSS{\footnote{KSS (named after the scientists who discovered it, Kovtun-Son-Starinets)
is a lower bound on the fluidity of the medium which is the ratio of shear viscosity to
entropy density and is found to be equals to $1/{4\pi}$.}} limits of $\eta/s$,
it needs little larger values of $\tau$, shown by green dash line in Fig.~(\ref{fig:Tau_KSS}). The EPNJL curve sits
in between NJL and PNJL curves as expected from their $M(T)$ pattern in Fig.~(\ref{fig:mass2}).

In Sec.~(\ref{sec:causal}), we have discussed about the causal aspects of dissipation phenomena.
Dissipation current and forces are linked instantaneously by Eqs.~(\ref{ESeta}), (\ref{ESkappa}),
(\ref{ESel}), which means they are communicated via infinite diffusion velocity ($v_T\rightarrow\infty$)
and hence, causally disconnected. Through the Eq.~(\ref{IS}), the causal
connection between shear-channel force and current can be established by introducing finite
shear relaxation time $\tau_\pi$. The relaxation time $\tau$, discussed earlier, can be called
collisional relaxation time to distinguish it from shear relaxation time $\tau_\pi$. 
To zoom in their differences, one can think $\tau$ as microscopic time scale, which is originated from
microscopic collision, while $\tau_\pi$ can be considered as macroscopic time scale, required
to satisfy causality. In practice, we take $\tau_\pi\approx\tau$ but actually they are different
time scale, which is pointed out in Ref.~\cite{Muronga}. 
A rigorous relation in RTA~\cite{Denicol} can connect them by
relation:
\bea
\tau_\pi &=&\tau \Big(\frac{u^\mu k_\mu}{T}\Big)^\lambda
\nn\\
&=&\tau \Big(\frac{E}{T}\Big)^\lambda ({\rm in~fluid~rest~frame})~,
\eea
where $\lambda$ is unknown parameter.
So the inequality, given in Eq.~(\ref{tpi_eta}), will get a rigorous
form:
\bea
\tau_\pi &>& \frac{\eta}{(\epsilon + P)}
\nn\\
\tau \Big(\frac{E}{T}\Big)^\lambda &>& \frac{\beta_\pi\tau}{(\epsilon + P)}
\nn\\
\Rightarrow v_T^{max}&=&\sqrt{\frac{\beta_\pi}{(\epsilon + P)} \Big(\frac{T}{E}\Big)^\lambda} < 1~,
\eea
where $\eta=\beta_\pi\tau$ is assumed. In general, we consider $\lambda=0$, which means $\tau_\pi=\tau$, i.e. the inequality becomes
\be
v_T^{max}=\sqrt{\frac{\beta_\pi}{(\epsilon + P)}} < 1
\ee

The maximum value of diffusion speed,
from Eq.~(\ref{vmax_tpi}), can be written for massless case as
\be
v_T^{max} \equiv \sqrt{\frac{\tau}{5\tau_\pi}}~,
\label{v_m0}
\ee
since $\eta=\tau(\epsilon +P)/5$ for massless fermionic/bosonic medium.
Now, one can easily recognize that $\tau_\pi\rightarrow0$ in Eq.~(\ref{vmax_tpi}) or
(\ref{v_m0}) give us $v_T^{max}\rightarrow\infty$. Hence to mention relativistic
inequality $v_T^{max}\leq 1$, massless matter should follow the inequality:
\be
\tau_\pi\geq \frac{\tau}{5}~.
\ee
This lower limit of $\tau_\pi$ ($=\tau/5$) is drawn by long dash 
line in Fig.~\ref{fig:Tau_KSS}(a). So in principle, $\tau_\pi$ can be
lower or greater than $\tau$ but we can bound it within the inequality:
$\frac{\tau}{5}\leq \tau_\pi \leq 10$ fm, where upper limit has been fixed 
from phenomenological side, by assuming $10$ fm life time of medium (shown by
straight horizontal long dash line). For $\tau_\pi\approx\tau$ approximation,
$v_T^{max}=1/\sqrt{5}$ massless case and $v_T^{max}(T)$ for NJL, PNJL and EPNJL
models are drawn in Fig.~\ref{fig:Tau_KSS}(b), where all curves follow $v_T^{max}\leq 1/\sqrt{5}$,
since we assume $\tau_\pi\approx \tau$.
However, we should accept that the general form of $v_T^{max}$ (at $\mu=0$):
\be
v_T^{max} =\sqrt{\Big(\frac{T}{E}\Big)^\lambda
\Big[\frac{\frac{\beta}{15}
	\int \frac{d^3\bp}{(2\pi)^3}\Big(\frac{\bp^2}{E}\Big)^2
	{f_Q}  (1-f_Q)}{\int \frac{d^3\bp}{(2\pi)^3}\Big(\frac{\bp^2}{3E}+E\Big){f_Q}}\Big]}~,
\ee	
whose massless limit should be
\be
\lim_{m\rightarrow 0}=\sqrt{\Big(\frac{1}{3}\Big)^\lambda\Big[\frac{1}{5}\Big]}~,
\ee
where roughly average energy can be considered as $E\approx 3T$.
Here, we have generated our numerical values for $\lambda=0$ i.e. for $\tau_\pi=\tau$ instead of going any general form.
At high $T$, all are merged to massless limit as expected and at low $T$, the values
of $v_T^{max}(T)$ are quite lower. It means that at low $T$ domain, 
$\tau_\pi$ is quite larger i.e. quite safer zone for causal aspects.
The inequality $\frac{\tau}{5}\leq \tau_\pi \leq 10$ fm is shown by arrow
in Fig.~\ref{fig:Tau_KSS}(a), where corresponding approximated values 
of $v_T^{max}$ are displayed in different zones. Here also, we have put massless ($M=0$) and confinement mass ($M=0.300$ GeV) curves (two solid black line) for $\tau_\pi=\tau$ in Fig.~\ref{fig:Tau_KSS}(a), (b).

\subsection{Finite $\mu$ Results}
%
 %
  \begin{figure} [ht]
  \includegraphics[scale=0.8]{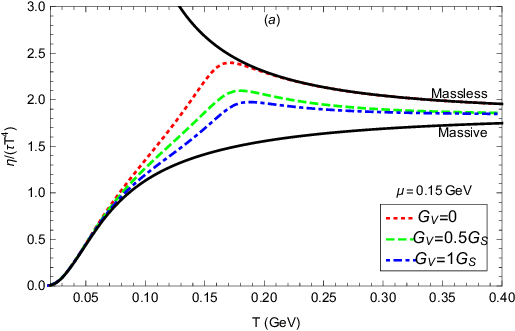}
  \includegraphics[scale=0.8]{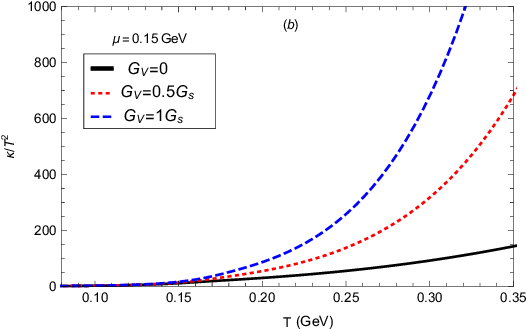}
  \includegraphics[scale=0.8]{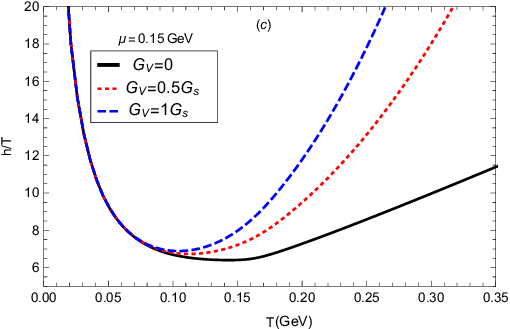}
  \caption{Finite density plots at different values of $G_V$ for (a) normalized shear viscosity, (b) thermal conductivity and (c) heat function as a function of $T$, in NJL model.}
  \label{fig:El_cond_gv}
  \end{figure}
Now, let us move to finite $\mu$ results, where we can explore the estimation of
thermal conductivity $\kappa$, which can not 
be studied in $\mu=0$ picture. However thermal diffusion coefficients can be estimated at $\mu=0$ (see in Ref.~\cite{Muronga}).
We also explore the effect of vector interaction, the incorporation of which becomes almost indispensable at nonzero $\mu$. In Fig.~\ref{fig:El_cond_gv}(a), we have plotted
$\eta/(\tau T^4)$ against $T$-axis for $\mu=0.150$ GeV. For massless case, instead
of a horizontal line, as obtained in Fig.~\ref{fig:ShT4}(a), we are getting a decreasing
function of temperature, shown by black solid line in Fig.~\ref{fig:El_cond_gv}(a). To understand the blowing trend in low temperature range, let us see an analytic form of $T$-dependence for finite $\mu$ by taking some rough assumption, described below.

For $\mu=0$, massless results of Eq.~(\ref{eta}) is 
\be
\frac{\eta}{\tau T^4}=
\Big(\frac{7}{8}\Big)\left(\frac{\pi^4}{90}\right)\Big(\frac{4\times (4N_FN_c)}{5\pi^2}\Big)\approx 1.84~,
\ee
which can be approximated as
\be
\frac{\eta}{\tau T^4}=\Big(\frac{4\times (4N_FN_c)}{5\pi^2}\Big)\approx 1.94~,
\label{eta_MB}
\ee
if we take Maxwell-Boltzmann distribution in place of Fermi-Dirac distribution of quark.
Now, for $\mu\neq 0$, Eq.~(\ref{eta_MB}) can get a simplified form
\bea
\eta &\approx& 2N_fN_c\frac{\beta}{15}
	\int \frac{d^3\bp}{(2\pi)^3}\tau\Big(\frac{\bp^2}{E}\Big)^2
	\Big[e^{-\beta(E-\mu)}
	\nn\\
	&&~~~~~~~~~~~~~~~~+e^{-\beta(E+\mu)}\Big]
\nn\\
	&=& 2N_fN_c\frac{\beta}{15}
	\int \frac{d^3\bp}{(2\pi)^3}\tau\bp^2e^{-\beta E}
	\Big[e^{\beta\mu}+e^{-\beta\mu}\Big]
\nn\\	
	&=&2N_FN_c\frac{4\tau T^4}{5\pi^2}\Big[e^{\beta\mu}+e^{-\beta\mu}\Big]
\nn\\
\Rightarrow\frac{\eta}{\tau T^4}&=&1.94 \times{\rm cosh}(\mu/T)~,
\eea
which can explain the blowing up nature of black solid line in Fig.~\ref{fig:El_cond_gv}(a),
when we decrease the temperature. Now if we revisit again Fig.~\ref{fig:ShT4}(a),
then we see that transition from $m=0$ to non-zero $M(T)$ provide a large suppression
at low $T$ domain, which is realized as the effect of non-perturbative QCD interaction.
Hence, in $\mu\neq 0$ picture, the transition of $m=0\rightarrow M(T)$ makes the blowing
up (black solid) curve be transformed to (red dotted) suppressed curve. Due to this turning,
we will get a peak-like behaviour around $T=160$ MeV. Now we know that with the increase of $\mu$, 
the transition temperature ($T_\Sigma$) decreases. Similarly, transition points
for transport coefficients like $\eta/(\tau T^4)$, $\sigma/(\tau T^2)$ as well as thermodynamical
quantities are also noticed to be shifted towards lower temperature as $\mu$ increases.
%
%
%

Now let us come to the vector interaction picture of NJL model.
As we introduce the vector interaction the 
transition temperature gets modified for a given chemical potential \textendash\, it starts 
increasing with the strength of $G_V$, which basically couples to the chemical potential 
through the relation $\tilde\mu=\mu-G_V n$, $\tilde\mu$ being the effective chemical potential.
It means that if we increase the value of $G_V$ the value of effective chemical potential decreases,
thus the transition temperature increases. Similar transition in the peak-like appearance of $\eta/(\tau T^4)$ 
is observed and it shifts towards higher $T$ as $G_V$ is increased. Apart from transition points, Fig.~\ref{fig:El_cond_gv}(a) also shows a decreasing profile for increasing
$G_V$.
The reason for the reduction of transport coefficient with vector interaction can be realized as follows.
We have already seen in Sec.~\ref{ssec:njl} and in Fig.~(\ref{fig:mass}), the constituent mass $M$
is slightly enhanced with $G_V$ near the transition temperature. On the other hand effective
chemical potential $\tilde{\mu}$ decreases with $G_V$. These increasing $M$ and decreasing 
$\tilde{\mu}$ make thermodynamical phase space part of $\eta$ reduce.
%


Similar to shear viscosity, electrical conductivity follow same pattern (not shown)
but totally different variation can be found for thermal conductivity,
as shown in Fig.~\ref{fig:El_cond_gv}(b). For thermal conductivity, heat function $h$, or more
precisely enthalpy density per net baryon/quark density plays an important role. Its temperature
dependence is shown in Fig.~\ref{fig:El_cond_gv}(c), where we see that $h$ increases with $G_V$
at high temperature, which dominantly appears in $\kappa$. Now, the reason for increasing 
$h$ with $G_V$ can be understood as follows. Increasing of $G_V$ make $\tilde{\mu}$ decrease
and so, $\rho$ decreases. Hence $h\propto 1/\rho$ increases.

\section{Summary} 
\label{Sum}
Present article has attempted to explore the effect of different extended components
of mean field models on transport coefficients calculations and 
made a comparison among them. 
First we start with NJL model which can map chiral phase transition of QCD medium. 
Here, quark condensate melts down near chiral transition temperature, around which
normalized transport coefficients and thermodynamical quantity like entropy density
also face maximum changes. While, bulk viscosity is showing peak structure
near transition temperature like interaction measure of QCD thermodynamics,
observed in LQCD and effective QCD model calculations. 
Hence, they may be roughly considered as alternative quantities
for mapping chiral phase transition.
Then to mimic QCD further closely we incorporate the deconfinement dynamics, 
along with the chiral one, by taking into account the background gauge field 
through PNJL model. Along with the chiral transition temperature, one can separately
identify deconfinement temperature, where Polyakov loop face a rapid change. 
The transport coefficients along with thermodynamical quantities will exhibit quite
interesting profile as they contain both chiral and deconfinement dynamics. Hence,
they show their signal like maximum change (shear viscosity, electrical conductivity) or maximum value (bulk viscosity) around an intermediate temperature between
chiral and deconfinement transition temperature.
After PNJL model, we have considered EPNJL model, which incorporate a strong entanglement between the chiral and 
deconfinement dynamics to enforce the coincidence of chiral and deconfinement transition
temperatures within the range provided by the LQCD. Due to this merging of two transitions temperatures,
we notice that order parameters (quark condensate, Polyakov loop), normalized-thermodynamical
quantities like entropy density, normalized-transport coefficients like shear viscosity, electrical conductivity 
are showing their maximum change or transition near same temperature. Bulk viscosity will show peak at that temperature. Massless case and NJL model
calculations of transport coefficients are coincided at high temperature, but PNJL and EPNJL
results still remain suppressed at high temperature due to transformation of Fermi-Dirac to modified
distribution function.

After exploring the thermodynamical phase-space components of transport coefficients, we have tried
to estimate relaxation time of quarks from the phenomenological understanding, which expect that 
shear viscosity to entropy density ratio will be very close to KSS bound. Imposing that phenomenological
expectation, required relaxation time from massless case to NJL to EPNJL to PNJL model become larger at low temperature,
but they merge at high temperature. Defining a shear relaxation time to satisfy causal aspect in the fluid, we have
shown its possible range for RHIC/LHC matter. In normal practice, (macroscopic) shear relaxation time is considered to be
equal with the (microscopic) particle relaxation time, but in reality the former time scale might be larger or smaller than the latter one. 
It is causality, which dictates that the shear diffusion speed in medium should not exceed the speed of light, for which we
get the lower limit of shear relaxation time. On the other hand, medium life time might be consider as upper limit of shear
relaxation time. Since shear diffusion speed from massless case to model calculations faces large suppression at low temperature,
therefore, we can say that non-perturbative low temperature zone of QCD is causally more safer zone. 

At the end, we have studied the finite quark chemical potential zone of quark matter. Similar features of decreasing transition temperature with increasing chemical potential is reflected through the appropriate shift of a peak-like appearance of normalized transport coefficients to a lower value of temperature. Role of vector interaction in NJL model and estimation of thermal conductivity
at finite quark chemical potential are also investigated.


{\bf Acknowledgment:} CAI would like to thank his institute, Tata Institute of 
Fundamental Research (India) funded by Department of Atomic Energy (DAE), Govt. of India. He would also like to acknowledge the University of Chinese Academy of Sciences, China for financial support. SG and JD acknowledge IIT Bhilai, funded by Ministry of Human
Resource Development (MHRD), Govt. of India. Authors are thankful to Amaresh Jashwal
for useful discussion.
\section{Appendix}
\subsection{LHS of RBE}
\label{sec:App1}
Here, we will address the detail calculation on left hand side (LHS) of RBE, 
as given in Eq.~(\ref{RBE2}) and see how it can be converted to different
(thermodynamical) gradient tensors, associated with shear viscosity, bulk viscosity, thermal conductivity,
electrical conductivity. Reader can find corresponding calculations of shear viscosity, bulk viscosity
parts from Refs.~\cite{Chakraborty:2010fr,Gavin:1985ph,Deb}, thermal conductivity part
from Refs.~\cite{Gavin:1985ph,Deb} and electrical conductivity part from 
Refs.~\cite{Greco,Kadam_el} separately but here we present in a combined form.
The FD distribution function, given in Eq.~(\ref{f_FD}), depends on macroscopic
quantities or fluid-element quantities - temperature $T$, chemical potential $\mu$
and four velocity $u^\mu$, which can depend on $x$ in local equilibrium picture.
It also depend on microscopic quantity or particle quantity - four momentum $p^\mu$,
which will not have any $x$ dependence. So Eq.~(\ref{f_FD}) can be re-written in local
equilibrium picture as
\be
f^0_{Q,\bar{Q}}=1/\{{\rm exp}\Big(\frac{p^\nu u_\nu(x)\mp\mu(x)}{T(x)}\Big)+1\}~.
\label{f_FDx}
\ee
Using Eq.~(\ref{f_FDx}) in first term of LHS of Eq.~(\ref{RBE2}), we get
\begin{widetext}
 \bea
\frac{p^\mu}{E} \del_{\mu} f^0_{Q,\bar{Q}}
&=& -\frac{{\rm exp}\Big(\frac{p^\nu u_\nu(x)\mp\mu(x)}{T(x)}\Big)}
{\{{\rm exp}\Big(\frac{p^\nu u_\nu(x)\mp\mu(x)}{T(x)}\Big)+1\}^2}
\Big[\frac{p^\mu}{E} \del_{\mu}\Big\{\frac{p^\nu u_\nu(x)\mp\mu(x)}{T(x)}\Big\}\Big]
\nn\\
&=& -f^0_{Q,\bar{Q}}(1-f^0_{Q,\bar{Q}})
\Big[\frac{p^\mu p^\nu}{ET(x)} \del_{\mu}u_\nu(x) -\frac{p^\mu p^\nu u_\nu}{ET^2(x)}\del_{\mu}T(x)
\mp \frac{p^\mu}{E}\del_\mu\Big(\frac{\mu(x)}{T(x)}\Big)\Big]~,
\nn\\
&=&-f^0_{Q,\bar{Q}}(1-f^0_{Q,\bar{Q}})\Big[\frac{p^\mu p^\nu}{ET}
\Big\{ \del_{\mu}u_\nu(x) -\frac{u_\nu\del_\mu T(x)}{T}\Big\}
+ \frac{2}{ET}\frac{dM_Q^2}{dT} u^\mu \partial_\mu T
\mp \frac{p^\mu}{E}\del_\mu\Big(\frac{\mu(x)}{T(x)}\Big)\Big]
\label{RBE3}
\eea
Our goal will be to express Eq.~(\ref{RBE3}) in terms of thermodynamical tensor ${\cal U^{\mu\nu}_\eta}$, $\del_\rho u^\rho$, ${\cal U^{\mu\nu}_\kappa}$, connected with $\eta$, $\zeta$, $\kappa$. 
Using the identity~\cite{Chakraborty:2010fr},
\be
\frac{\del_\mu T}{T}=u^\alpha\del_\alpha u_\mu - c_s^2 u_\mu\del_\alpha u^\alpha~,
\ee
with square of speed of sound $c_s^2=\Big(\frac{\del P}{\del \ep}\Big)$, we can get 
\bea
\frac{p^\mu}{E} \del_{\mu} f^0_{Q,\bar{Q}}
&=&-f^0_{Q,\bar{Q}}(1-f^0_{Q,\bar{Q}})\Bigg[\frac{p^\mu p^\nu}{2ET}
\Big\{ D_\mu u_\nu +D_\nu u_\mu-\frac{2}{3}\Delta_{\mu\nu}\del_\alpha u^\alpha \Big\}
- \frac{1}{3ET}\Bigg\{p^\mu p^\nu (\Delta_\munu +3 c_s^2 u_\mu u_\nu)
\nn\\
&& ~~~~~~~~~~~~~~~~~~~~~~~~ -3 c_s^2 T^2 \frac{dM_Q^2}{dT}\Bigg\} (\del_\rho u^\rho)
\mp \frac{p^\mu}{E}\del_\mu\Big(\frac{\mu(x)}{T(x)}\Big)\Bigg]
.
\label{U_zeta}
\eea
Using Eq.~(\ref{U_zeta}) and (\ref{RBE2}), we get 
\bea
-f^0_{Q,\bar{Q}}(1-f^0_{Q,\bar{Q}}) \left[\frac{p^\mu p^\nu}{2ET}{\cal U^{\mu\nu}_\eta} 
	- \frac{1}{3ET}\Bigg\{p^\mu p^\nu (\Delta_\munu +3 c_s^2 u_\mu u_\nu)
	 -3 c_s^2 T^2 \frac{dM_Q^2}{dT}\Bigg\} (\del_\rho u^\rho)
	\right]&=&
-\Big(\frac{p\cdot u}{E T}\Big)\frac{1}{\tau_{Q, \bar{Q}}}\Big[A_{\mu\nu}^{\FB{Q,\bar{Q}}} {\cal U}^{\mu\nu}_\eta 
\nn\\ 
	&&+ Z^{\FB{Q,\bar{Q}}} (\partial_\rho u^\rho)
		\Big]f^0_{Q,\bar{Q}}(1-f^0_{Q,\bar{Q}}).
	\nn	\\
\eea
Now, from the above equation comparing the coefficients of $\cal U^{\mu\nu}_\eta$ and $\del_\rho u^\rho$ in both side~\cite{Chakraborty:2010fr}, 
\bea
\Rightarrow A_{\mu\nu}^{\FB{Q,\bar{Q}}}&=&\tau_{Q,\bar{Q}}\frac{ \: p_\mu p_\nu}{2E} 
~~~({\rm using}~p\cdot u=E)
\\
{\rm and,~~}
Z^{\FB{Q,\bar{Q}}} &=& -\frac{\tau_{Q, \bar{Q}}}{3E} \left\{p^\mu p^\nu (\Delta_\munu +3 c_s^2 u_\mu u_\nu)
-3 c_s^2 T^2 \frac{dM_Q^2}{dT} \right\}
\nn\\
&=& \tau_{Q,\bar{Q}}
\frac{1}{3E} \left[\bp^2- 3c_s^2\left(E^2-T^2\frac{dM_Q^2}{dT^2}\right)\right]~~~ (\rm in~local~rest~frame.)~.
\label{bulkz}
\eea
The solution (Eq. \ref{bulkz} ) is not unique~\cite{Chakraborty:2010fr}. One can make a shift 
$Z^{\FB{Q,\bar{Q}}} \rightarrow Z^{\prime\FB{Q,\bar{Q}}} 
= Z^{\FB{Q,\bar{Q}}} - a - bE$, which can also be true.
The unknown constants $a,b$ are associated with particle
number and energy conservation respectively. 
Here we calculate bulk viscosity for zero chemical potential 
($\mu = 0$), thus $a=0$.
Now, if we have a particular solution of Eq. (\ref{bulkz}) 
as $Z_{\rm par}^{\FB{Q,\bar{Q}}}$ which satisfy Landau-Lifshitz condition
(fluid frame is at rest with energy flow) then
$Z^{\FB{Q,\bar{Q}}}=Z_{\rm par}^{\FB{Q,\bar{Q}}}- bE$.
With the help of microscopic definition of thermodynamical quantities 
such as entropy density ($s$), heat capacity $c_V$ and 
Eq. {\ref{Pi}} we can find the bulk pressure as
\be
  \Pi = 2N_FN_c\beta
  \int \frac{d^3\bp}{(2\pi)^3}
  \frac{Z_{\rm par}^{\FB{Q,\bar{Q}}}}{3E} \left[\bp^2- 3c_s^2\left(E^2-T^2\frac{dM_Q^2}{dT^2}\right)\right]
  \Big\{{f^0_Q}  (1-f^0_Q)
  + f^0_{\bar Q} (1-f^0_{\bar Q})\Big\}(\partial_\rho u^\rho)~ 
\ee
and,
\be
Z_{\rm par}^{\FB{Q,\bar{Q}}} = \tau_{Q,\bar{Q}}
\frac{1}{3E} \left[\bp^2- 3c_s^2\left(E^2-T^2\frac{dM_Q^2}{dT^2}\right)\right]~.
\ee
Here, we can express square of speed of sound at $\mu=0$ as $c_s^2=\frac{s}{c_V}=\frac{s}{T\Big(\frac{ds}{dT}\Big)_V}$.

Now, Eq.~(\ref{RBE3}) also carry
${\cal U}^\mu_\kappa$, related with $\kappa$, which can be constructed by combining last
two terms of Eq.~(\ref{RBE3}):
\bea
-\frac{p^\mu (p\cdot u)\del_\mu T(x)}{ET^2}
\mp \frac{p^\mu}{E}\del_\mu\Big(\frac{\mu(x)}{T(x)}\Big)
&=&\frac{p^\mu}{E}\Big\{-\frac{(p\cdot u)\del_\mu T(x)}{T^2}
\mp \del_\mu\Big(\frac{\mu(x)}{T(x)}\Big)\Big\}
\nn\\
&\equiv&\frac{p^i}{E}\Big\{-\frac{E\del_i T(x)}{T^2}
\mp \del_i\Big(\frac{\mu(x)}{T(x)}\Big)\Big\} ~~~({\rm using}~p\cdot u=E~{\rm and}~p^\mu\equiv p^i)
\nn\\
&=&\frac{p^i}{E}\Big\{\frac{E}{h} \mp 1\Big\}\del_i\Big(\frac{\mu(x)}{T(x)}\Big) 
~~~(Eq.~(\ref{GD}) ~{\rm with}~\del_i P=0~{\rm is}~-\frac{\del_i T}{T^2}=\frac{1}{h}\del_i(\mu/T) )
\nn\\
\label{U_kappa}
\eea
Using Eq.~(\ref{U_kappa}) in (\ref{RBE3}) and then in (\ref{RBE2}), we get 
\bea
-f^0_{Q,\bar{Q}}(1-f^0_{Q,\bar{Q}})\frac{p^i}{E}\Big\{\frac{E}{h} \mp 1\Big\}\del_i\Big(\frac{\mu(x)}{T(x)}\Big) +...&=&
-\Big(\frac{p\cdot u}{E T}\Big)\frac{1}{\tau_{Q, \bar{Q}}}\Big[B^i\frac{T^2}{h}\del_i\Big(\frac{\mu(x)}{T(x)}\Big) +...
\Big]f^0_{Q,\bar{Q}}(1-f^0_{Q,\bar{Q}}
\nn\\
\Rightarrow B^{i}_{\FB{Q,\bar{Q}}}&=&\tau_{Q,\bar{Q}}\frac{ \: p_i}{E T}(E \mp h) 
~~~({\rm using}~p\cdot u=E)
\nn\\
\Rightarrow B^{\mu}_{\FB{Q,\bar{Q}}}&=&\tau_{Q,\bar{Q}}\frac{ \: p_\mu}{E T}(E \mp h)
\eea

On the other hand, the second term of LHS of Eq.~(\ref{RBE2}) can be simplified 
through 4-vector to 3-vector and again to 4-vector components (i.e.
$\mu\rightarrow i\rightarrow \mu$ index) as
\bea
e_{Q,\bar{Q}}F^{\mu\nu}\frac{p_\nu}{E}\frac{\del f^0_{Q,\bar{Q}}}{\del p^\mu}
&\equiv&
e_{Q,\bar{Q}}E^i\frac{\del f^0_{Q,\bar{Q}}}{\del p^i}
\nn\\
&=&-f^0_{Q,\bar{Q}}(1-f^0_{Q,\bar{Q}})\Big[e_{Q,\bar{Q}}E^i\frac{\del}{\del p^i}\Big(\frac{E}{T}\Big)\Big]
\nn\\
&=&-f^0_{Q,\bar{Q}}(1-f^0_{Q,\bar{Q}})\Big[e_{Q,\bar{Q}}\frac{{\vec E}\cdot{\vec p}}{ET}\Big]
\nn\\
&=&-f^0_{Q,\bar{Q}}(1-f^0_{Q,\bar{Q}})\Big[e_{Q,\bar{Q}}\frac{E_\mu p^\mu}{ET}\Big]~.
\label{U_el}
\eea
Since electromagnetic field tensor $F^{\mu\nu}$ carries only electric fields (as no external magnetic field is considered in the present work), so $F^{\mu\nu}\frac{p_\nu}{E}=F^{0\mu}+F^{ij}\frac{p_j}{E}+..=F^{0\mu}=E^\mu$ is used in the above calculations.

Using Eq.~(\ref{U_el}) in (\ref{RBE3}) and then in (\ref{RBE2}), we get
\bea
-f^0_{Q,\bar{Q}}(1-f^0_{Q,\bar{Q}})\Big[e_{Q,\bar{Q}}\frac{E_\mu p^\mu}{ET}\Big] +...&=&
-\Big(\frac{p\cdot u}{E T}\Big)\frac{1}{\tau_{Q, \bar{Q}}}\Big[C_{\mu}^{\FB{Q,\bar{Q}}} {E}^{\mu}+...
\Big]f^0_{Q,\bar{Q}}(1-f^0_{Q,\bar{Q}}
\nn\\
\Rightarrow C^\mu&=&\tau_{Q,\bar{Q}}\frac{e_{Q, {\bar Q}} \: p^\mu}{E}~.
\eea

\subsection{PNJL/EPNJL distribution replacement}
\label{sec:App2}
The modified distribution function, given in Eq.~(\ref{PNJL_f}), is
can be realized as color average of FD distribution of color particle
with imaginary chemical potential~\cite{HMPQM1}. Let us write down the 
FD distribution with imaginary chemical potential $Q^i$ in local equilibrium picture 
as~\cite{Hidaka}
\be
f^i_{Q,\bar{Q}}=1/\{{\rm exp}\Big(\frac{p^\nu u_\nu(x)\mp\mu(x) \mp Q^i}{T(x)}\Big)+1\}~,
\label{f_PNJLx}
\ee
where $Q^i=2\pi T (+q, 0, -q)$ with dimensionless condensate variable $q$. 
The Polyakov loop variable can be expressed as
\bea
\Phi &=& \frac{1}{3}\sum_i e^{i\beta Q_i}
\nn\\
&=&\frac{1}{3}\{1+2cos(2\pi q)\}~.
\label{Phi_q}
\eea
Let us rename the modified distribution function as $f_\Phi$ and
rewrite as
\be
f^\Phi_{Q,\bar{Q}}=\frac{\Phi e^{-\beta(E\mp\mu)}
 +2\bar{\Phi}e^{-2\beta(E\mp\mu)}+e^{-3\beta(E\mp\mu)}}
 {1+3\Phi e^{-\beta(E\mp\mu)}+3\bar{\Phi}e^{-2\beta(E\mp\mu)}~.
 +e^{-3\beta(E\mp\mu)}}=\frac{N}{D} (say)
\ee 
One can easily check the relation between $f^i_{Q,\bar{Q}}$ and $f^\Phi_{Q,\bar{Q}}$
as
\bea
\frac{1}{3}\sum_i f^i_{Q,\bar{Q}}
&=& \frac{1}{3}\Big[\frac{1}{{\rm exp}\{(E\mp \mu \mp i2\pi Tq)/T\}+1} 
+\frac{1}{{\rm exp}\{(E\mp \mu)/T\}+1} +\frac{1}{{\rm exp}\{(E\mp \mu \pm i2\pi Tq)/T\}+1}\Big]
\nn\\
&=&\frac{1}{3}\Big[\frac{e^{2\beta(E\mp\mu)}\{1+2cos(2\pi q)\}+2e^{\beta(E\mp\mu)}\{1+2cos(2\pi q)\}+3}
{e^{3\beta(E\mp\mu)}+e^{2\beta(E\mp\mu)}\{1+2cos(2\pi q)\}+e^{\beta(E\mp\mu)}\{1+2cos(2\pi q)\}+1}\Big]
\nn\\
&=&\Big[\frac{e^{2\beta(E\mp\mu)}\Phi+2e^{\beta(E\mp\mu)}\Phi +1}
{e^{3\beta(E\mp\mu)}+3e^{2\beta(E\mp\mu)}\Phi +3e^{\beta(E\mp\mu)}\Phi +1}\Big]~~~~({\rm using~ Eq.}~(\ref{Phi_q}))
\nn\\
&=&f^\Phi_{Q,\bar{Q}}
\eea
The transport coefficient calculations remain almost same only the terms,
associated with distribution will have to be recalculate. When we start
our journey from color particle FD distribution $f^i_{Q,\bar{Q}}$, and its 
color average $\frac{1}{3}\sum f^i_{Q,\bar{Q}}$, then their 
derivative with respect to $E$, $p_i$ and $x$ will have same anatomy as earlier
i.e.
\bea
\frac{1}{3}\sum_i\frac{\del f^i_{Q,\bar{Q}}}{\del E}&=&-\beta \frac{1}{3}\sum_if^i_{Q,\bar{Q}}(1-f^i_{Q,\bar{Q}})~,
\nn\\
\frac{1}{3}\sum_i\frac{\del f^i_{Q,\bar{Q}}}{\del p_i}
&=&-\beta \frac{1}{3}\sum_i\Big(\frac{p^i}{E}\Big) f^i_{Q,\bar{Q}}(1-f^i_{Q,\bar{Q}})~,
\nn\\
\frac{1}{3}\sum_i\del_\mu f^i_{Q,\bar{Q}}&=&-\frac{1}{3}\sum_i f^i_{Q,\bar{Q}}(1-f^i_{Q,\bar{Q}})
\del_{\mu}\Big\{\frac{p^\nu u_\nu(x)\mp\mu(x)}{T(x)}\Big\}
\label{fi_fPhi}
\eea
These relations indicate that anatomy of Eqs.~(\ref{df_phi}), (\ref{RBE2}), (\ref{RBE3})
remain same. Only FD distribution is replaced by FD distribution of color particles.
Now we have to transform FD distribution of color particles $f^i_{Q,\bar{Q}}$ in Eq.~(\ref{fi_fPhi}) 
into modified distribution function $f^\Phi_{Q,\bar{Q}}$. We can express the terms
$f^i_{Q,\bar{Q}}(1-f^i_{Q,\bar{Q}})$ as
\bea
-\frac{1}{3}\sum_if^i_{Q,\bar{Q}}(1-f^i_{Q,\bar{Q}}) &=&
\frac{1}{3\beta}\sum_i\frac{\del f^i_{Q,\bar{Q}}}{\del E}
\nn\\
&=& \frac{1}{\beta}\frac{\del f^\Phi_{Q,\bar{Q}}}{\del E}
\nn\\
&=& \frac{1}{\beta}\frac{\Big(D\frac{\del N}{\del E}-N\frac{\del D}{\del E}\Big)}{D^2}
\eea
If we expand the above expression, we can get 
\bea
\frac{\Big(D\frac{\del N}{\del E}-N\frac{\del D}{\del E}\Big)}{D^2}
&=&\frac{N}{D}\Big(1-\frac{N}{D}\Big) + \{2D(\Phi e^{-2\beta(E\mp\mu)}+e^{-3\beta(E\mp\mu)})-2N^2 \}/D^2
\nn\\
&=&f^\Phi_{Q,\bar{Q}}(1-f^\Phi_{Q,\bar{Q}}) 
+ 2e^{-2\beta(E\mp\mu)}[(1+e^{-2\beta(E\mp\mu)})\Phi(1-\Phi)+e^{-\beta(E\mp\mu)}(1-\Phi^2)]/D^2
\nn\\
&\approx&f^\Phi_{Q,\bar{Q}}(1-f^\Phi_{Q,\bar{Q}}) ~.
\label{extra}
\eea
 \begin{figure} [ht]
 \includegraphics[scale=0.3]{pnjlcheck.eps}
 \caption{The ratio between phase-space integration with approximated (excluding extra term) and exact (including extra term) vs $T$, whose values close to 1 reflects that one may go with this approximation.}
 \label{fig:check}
 \end{figure}
The extra term in Eq.~(\ref{extra}) might be ignored with respect to the dominating term $f^\Phi_{Q,\bar{Q}}(1-f^\Phi_{Q,\bar{Q}})$. For numerical check, Fig.~(\ref{fig:check}) has shown the ratio between phase-space integration with approximated (excluding extra term) and exact (including extra term) i.e.
\be
\frac{\chi_{\rm approx}}{\chi_{\rm exact}} =
\frac{\int \frac{d^3k}{(2\pi)^3}\Big[\beta f^\Phi_0(1-f_0^\Phi)\Big]}{\int \frac{d^3k}{(2\pi)^3}\Big[-\frac{1}{3}\sum_i\frac{\del f^i_{Q,\bar{Q}}}{\del E}\Big]}~.
\ee
We notice that the extra term roughly contribute upto $10\%$. So one may go safely for rough estimation of different transport coefficients with simplified phase-space factor $\Big[\beta f^\Phi_0(1-f_0^\Phi)\Big]$ instead of its complicated version $\Big[-\frac{1}{3}\sum_i\frac{\del f^i_{Q,\bar{Q}}}{\del E}\Big]$ or $\frac{\Big(D\frac{\del N}{\del E}-N\frac{\del D}{\del E}\Big)}{D^2}$. We have shown here susceptibility-type quantity $\chi$, which is basically attached with all transport coefficients, hence this approximation will be valid during estimation transport coefficients or any other quantities, which are proportionally connected with susceptibility. Owing to this assumption, we have used the replacement identity
\be
-\frac{1}{3}\sum_if^i_{Q,\bar{Q}}(1-f^i_{Q,\bar{Q}})\approx
-f^\Phi_{Q,\bar{Q}}(1-f^\Phi_{Q,\bar{Q}})
\ee
during the calculation of different transport coefficients in PNJL and EPNJL models.
\end{widetext}

\end{document}